\documentclass[11pt,rotate,epsf]{article}
\usepackage[utf8]{inputenc}

\usepackage{courier}        
\usepackage{graphicx}        

\usepackage{amsfonts}
\usepackage{amsthm, amssymb}
\usepackage[tbtags]{amsmath}
\usepackage{pdfpages}
\usepackage{subfig}
\usepackage{float}
\usepackage{bbm}
\usepackage{blkarray}
\usepackage{authblk}
\usepackage{natbib}
\usepackage{mathtools}
\usepackage{booktabs}
\usepackage{rotating}
\usepackage[margin=3.6cm]{geometry}
\usepackage{xcolor}
\usepackage[linesnumbered,ruled,lined,boxed]{algorithm2e}
\usepackage{hyperref}
\usepackage{multirow}
\usepackage[normalem]{ulem}
\usepackage{adjustbox}

\hypersetup{
    colorlinks,
    linkcolor={red!50!black},
    citecolor={green!50!black},
    urlcolor={blue!80!black}
}
\SetAlFnt{\small}
\SetAlCapFnt{\large}
\SetAlCapNameFnt{\large}

\definecolor{darkgreen}{rgb}{0.0, 0.5, 0.0}

\def\0{\mbox{\bf{0}}}

\newcommand\indsim{\stackrel{\mathclap{\normalfont\mbox{\textit{ind}}}}{\sim}}

\theoremstyle{definition}
\newtheorem{prop}{Proposition}

\graphicspath{{Figures/}}

\begin{document}

\title{\bf Informative co-data learning for high-dimensional Horseshoe regression}

\author{Claudio Busatto}
\affil{Department of Statistics, Computer Science, Applications “G. Parenti”, University of Florence, Florence, Italy}

\author{Mark van de Wiel}
\affil{Dep. Epidemiology \& Data Science, Amsterdam Public Health research institute, Amsterdam University Medical Centers, Amsterdam, The Netherlands}

\date{\vspace{-5pt}}

\maketitle
\begin{abstract}
\noindent High-dimensional data often arise from clinical genomics research to infer relevant predictors of a particular trait. A way to improve the predictive performance is to include information on the predictors derived from prior knowledge or previous studies. Such information is also referred to as ``co-data''. To this aim, we develop a novel Bayesian model for including co-data in a high-dimensional regression framework, called Informative Horseshoe regression (infHS). The proposed approach regresses the prior variances of the regression parameters on the co-data variables, improving variable selection and prediction. We implement both a Gibbs sampler and a Variational approximation algorithm. The former is suited for applications of moderate dimensions which, besides prediction, target posterior inference, whereas the computational efficiency of the latter allows handling a very large number of variables. We show the benefits from including co-data with a simulation study. Eventually, we demonstrate that infHS outperforms competing approaches for two genomics applications.
\end{abstract}


%

\section{Introduction}
\label{sec:introduction}
%
The analysis of high-dimensional data sets is a main interest in many scientific fields. In particular, clinical research often deals with huge amount of data, such as genes expression or genome-wide methylation levels for relatively few samples, due to budget or practical constraints. We consider regression models to study a clinical outcome with this type of data. Since the number of parameters, $p$, overwhelms sample size, $n$, we develop an approach to improve the overall performance of the model by incorporating (prior) external knowledge in the estimating process. Such an external source of information is referred to as co-data \citep*[complementary data;][]{neuenschwander2016use}, as it provides additional information about the covariates. We consider two different types: continuous, such as $p$-values from previous studies, or categorical, such as membership to a group, e.g. chromosome or particular genomic regions.

Several methods allow to incorporate one source of auxiliary information in a regression framework \citep*{Tai-2007,Boonstra-2013}. A popular method is the (sparse) group LASSO \citep{Yuan-2006, simon2013sparse}, which penalizes groups of variables using one common hyperparameter for all groups. Such a solution is attractive when the number of covariate groups is large, but lacks flexibility and fails to adapt locally in other settings, leading to sub-optimal results \citep{Munch-2019}. More recent work focuses on the estimation of adaptive penalties with prior variances specific for each group \citep*[][\citet{Munch-2019}]{vandeWiel-2019, Velten-2019}. These methods, however, are restrictive in use as they either deal with just one (discrete) co-data source or handle one specific type of outcome only. In \citet*{vanNee-2021}, instead, a Ridge regression method is proposed that allows for multiple co-data sources by regressing the local variances on the co-data. With their approach, the co-data regression parameters are estimated independently for each co-data source and are eventually combined using a vector of weights, where each weight is related to the importance of a single co-data source.

Here, we present a novel Bayesian method for both linear and binary regression that accounts for multiple co-data information. In particular, we introduce a generalization of the Horseshoe regression \citep*{Carvalho-2010}, referred to as the \textit{informative Horseshoe regression} model (infHS). We regress the local variances on the co-data variables, following the work of \citet{vanNee-2021}. In contrast to \citet{Kpogbezan-2019}, where the Horseshoe prior is used with only a single two-group co-data source, our model is flexible with respect to the co-data type, as it allows for both continuous and discrete co-data predictors. Moreover, it extends to binary outcome via probit regression \citep*{Albert-1993}. Unlike \citet{vanNee-2021}, it tackles the sparse setting and co-data regression parameters related to different sources are estimated jointly, avoiding multiple regressions for each co-data source separately.

We first propose a Gibbs sampler for iteratively updating posterior parameters. We introduce a novel rejection sampling method for sampling from the non-analytical full-conditional distribution of the local variances. When the number of variables increases, we rely on the computational methods presented in \citet*{Bhattacharya-2016} to sample efficiently from a multivariate normal density. To make the method applicable to particularly large $p$ settings, we develop a Variational Bayes approximation to the joint posterior distribution, using techniques from \citet{Munch-2019} to efficiently optimize the target density of the variational distribution, rendering an algorithm with computation time linear in $p$. With simulations and two data applications, we show that both prediction and variable selection benefit from the inclusion of co-data information, the latter being particularly relevant under the Horseshoe setting.

The paper in organized as follows. Section \ref{sec:model} introduces the hierarchical structure of the model and discusses the parametrizations. In Sections \ref{sec:posterior} and \ref{sec:VB} we develop the Gibbs sampling algorithm and the Variational Bayes approximation to the joint posterior distribution, respectively. Section \ref{sec:simulation} illustrates the benefit of co-data information on variable selection with a simulation study, whereas Section \ref{sec:realdata} presents applications of our model to two data sets, one from genetics and one from cancer genomics. We conclude with discussions and possible extensions in Section \ref{sec:conclusions}.

\section{The model}
\label{sec:model}
%

Let $\mathbf{y} \in \mathbb{R}^{n}$ be the response vector and $\mathbf{X} \in \mathbb{R}^{n \times (p+1)}$ the design matrix, with the first column of ones. We regress $\mathbf{y}$ on $\mathbf{X}$ using a generalized linear model (GLM) with regression coefficient vector $\boldsymbol{\beta} = \left[\beta_0 \; \beta_1 \dots \beta_p\right]^\intercal$.
\begin{align*}
	Y_i \mid \mathbf{x}_i, \boldsymbol{\beta} \; \indsim & \; p\left(Y_i \mid \mathbf{x}_i, \boldsymbol{\beta}\right) \\
	E_{Y_i \mid \mathbf{x}_i, \boldsymbol{\beta}}\left(Y_i\right) = & \; g^{-1} \left(\mathbf{x}_i^\intercal \boldsymbol{\beta}\right), \quad i = 1, \dots, n,
\end{align*}
where $\mathbf{x}_i = \left[1 \; x_{i1} \dots x_{ip} \right]^\intercal$ is the set of covariates related to observation $i$. 

Following \cite{Carvalho-2010}, the Horseshoe prior locally shrinks regression parameters toward zero and provides a sparse solution for $\boldsymbol{\beta}$. We assume a normal prior distribution for each $\beta_j$, where the variance is decomposed in a global scale parameter $\tau$ and a local shrinkage parameter $\lambda_j$. Formally,
\begin{align*}
    \beta_0 \mid \sigma^2, \tau, \lambda_0 \sim & \; \mathcal{N}\left(0, \sigma^2 \tau^2 \lambda_0^2 \right), \\
    \lambda_0 \sim &\; \mathcal{C}^+(0, 1), \\
    \beta_j \mid \sigma^2, \tau, \lambda_j \sim & \; \mathcal{N}\left(0, \sigma^2 \tau^2 \lambda_j^2 \right), \quad j = 1, \dots, p.
\end{align*}
Suppose that $D$ different co-data sources $\mathbf{Z}_d \in \mathbb{R}^{p \times m_d}$ are available, where $\sum_d m_d = M$ and $d = 1, \dots, D$. In order to capture the external information effect, \cite{vanNee-2021} introduce parameters $\omega_d > 0$ and $\boldsymbol{\gamma}_d \in \mathbb{R}^{m_d}$, $d = 1, \dots, D$, to model covariate-specific shrinkage $\lambda_j$, $j = 1, \dots, p$, as a function of $\mathbf{Z} = \left[\mathbf{Z}_1, \dots, \mathbf{Z}_D\right]$. Parameters $\boldsymbol{\gamma}_d$ represent the regression coefficient vector related to matrix $\mathbf{Z}_d$ and are estimated separately for each group $d$, whereas co-data weights $\omega_d > 0$ model the relative importance of group $d$ and are introduced to combine the different co-data sources. Here, parameter $\boldsymbol{\gamma} = \left[\boldsymbol{\gamma}_1^\intercal \dots \boldsymbol{\gamma}_D^\intercal \right]^\intercal$ is update jointly by sampling from its full-conditional distribution. This way the co-data sources are naturally combined and grouping weights $\omega_d$ can be excluded from the model. Therefore, the hierarchical set of prior distributions in our model is
\begin{equation}
\label{eq:priorlambdaj}
    \begin{aligned}	
        \lambda_j \mid \mathbf{Z}, \boldsymbol{\gamma} \sim & \; \mathcal{C} \left(\sum_{d = 1}^D \left(\mathbf{z}_j^d\right)^\intercal \boldsymbol{\gamma}_d, s_0^2 \right) \cdot \mathbb{I}_{\left(\lambda_j > 0\right)},  \quad j = 1, \dots, p,  \\
        \boldsymbol{\gamma}_d \mid \boldsymbol{\Sigma}_{\gamma_d} \sim & \; \mathcal{N}_{m_d} \left(\mathbf{0}, \boldsymbol{\Sigma}_{\gamma_d}\right), \quad d = 1, \dots, D,   \\
	\tau \sim & \; \mathcal{C}^+ \left(0, 1\right), \\
        \sigma^2 \sim &\; \mathcal{IG}\left(v, q\right), 
    \end{aligned}
\end{equation}
where $\mathcal{C}$ denotes the Cauchy distribution and $\mathcal{C}^+$ its half-positive part. Here we model the location parameter of the prior local variances mainly because it allows to sample co-data coefficients $\boldsymbol{\gamma}$ jointly from a multivariate normal distribution (See Section \ref{sec:posterior}). Note that when the co-data are not informative and $\boldsymbol{\gamma}_d \to 0$ for each $d$, the model reduces to the ordinary Horseshoe prior from \cite{Carvalho-2010} with $s_0^2 = 1$.

In \cite{vandeWiel-2019} the authors estimate prior covariance matrices $\boldsymbol{\Sigma}_d$ from the data with an Empirical Bayes estimator separately for each source. This approach, however, is computationally burdensome, as it implies the implementation of multiple MCMC chains until convergence. Here we consider a group-specific scale parameter $\kappa_d^2$ and a ridge-like prior $\boldsymbol{\Sigma}_d = \kappa_d^2 \mathbf{I}_{m_d}$. 
A main advantage of this approach is the computational efficiency, because only prior scale parameters $\kappa_d^2$ have to be updated. A conjugated prior distribution for $\kappa_d^2$ is
\begin{align}
	\kappa_d^2 \sim & \; \mathcal{IG}\left(a_d, b_d \right). \label{eq:pkappa}
\end{align}
Parameters $\kappa_d^2$ act deep in the model and one can argue that they should have a small impact on the global estimation process. For this reason, a non-informative choice for $a_d$ and $b_d$ should suffice.

\section{Posterior inference}
\label{sec:posterior}

In this section we introduce a Gibbs sampler that iteratively updates the parameters by sampling from their full-conditional distributions. We show the details of the algorithm for the linear regression model, under the assumption
\begin{equation*}
	y_i = \mathbf{x}_i^\intercal \boldsymbol{\beta} + \varepsilon_i, \quad \varepsilon_i \sim \mathcal{N}\left(0, \sigma^2\right), \quad i = 1, \dots, n.
\end{equation*}
However, P\'olya-Gamma latent variables \citep*{Polson-2013} or probit GLM \citep{Albert-1993} can be introduced to augment the model and reach a gaussian full-conditional distribution for $\boldsymbol{\beta}$ in a binary regression model.

\vspace{0.4cm}
\noindent \textbf{\textit{1. Sampling $\boldsymbol{\beta}$ and $\sigma^2$.}}
The full-conditional distributions of $\boldsymbol{\beta}$ and $\sigma^2$ are
\begin{align}
	\boldsymbol{\beta} \mid \mathbf{y}, \mathbf{X}, \sigma^2, \tau^2, \lambda_0^2, \boldsymbol{\lambda} \sim & \;\mathcal{N}_{p+1}\left(\boldsymbol{\Sigma}_\beta^\star \mathbf{X}^\intercal \mathbf{y}, \sigma^2 \boldsymbol{\Sigma}_\beta^\star \right), \quad \boldsymbol{\Sigma}_\beta^\star = \left(\mathbf{X}^\intercal \mathbf{X} + \tau^{-2}\boldsymbol{\Lambda}^{-2}\right)^{-1}, \label{eq:postbeta} \\
	\sigma^2 \mid \mathbf{y}, \mathbf{X}, \boldsymbol{\beta}, \tau^2, \lambda_0^2, \boldsymbol{\lambda} \sim & \; \mathcal{IG}\left(v + \frac{n+p+1}{2}, q + \frac{1}{2}\Vert \mathbf{y} - \mathbf{X} \boldsymbol{\beta}\Vert_2^2 + \frac{\beta_0^2}{2 \tau^2\lambda_0^2} + \frac{1}{2\tau^2}\sum_{j = 1}^p \frac{\beta_j^2}{\lambda_j^2}\right), \nonumber
\end{align}
where $\boldsymbol{\Lambda} = \text{diag}\left(\lambda_0, \lambda_1, \dots, \lambda_p\right)$. Sampling from \eqref{eq:postbeta} requires the inversion of the $p \times p$ covariance matrix $\boldsymbol{\Sigma}_\beta^\star$, which becomes computationally infeasible when the number of covariates $p$ increases, as naive inversion is of order $\mathcal{O}\left(p^3\right)$. For high-dimensional problems we rely on the strategies introduced in \cite{Bhattacharya-2016} and \citet*{Johndrow-2020} to reduce the computational costs of sampling regression parameters $\boldsymbol{\beta}$ to $\mathcal{O}\left(n^2 p\right)$ operations.

\vspace{0.4cm}
\noindent \textbf{\textit{2. Sampling $\boldsymbol{\lambda}$, $\boldsymbol{\gamma}$ and $\boldsymbol{\kappa}^2$.}}
The prior distributions for $\lambda_j$, $j \ne 0$, and $\boldsymbol{\gamma}$ in \eqref{eq:priorlambdaj} are not conjugated. To this aim, we rely on the data-augmentation step proposed in \cite{Geweke-1993} to reach a conjugated framework and jointly update parameter $\boldsymbol{\gamma}$ by sampling from a multivariate normal distribution. 

\begin{prop}
	Let $x$ and $y$ be random variables such that 
	\begin{equation*}
		x \sim \mathcal{C}\left(m, s^2\right) \quad \text{and} \quad y \sim \mathsf{t}_v\left(m, s^2\right),
	\end{equation*}
	where $t_v$ denotes the Student-$t$ distribution. If $v = 1$, then $X \sim Y$. 
	\label{prop1}
\end{prop}

\begin{prop}
	Let $x$ and $\theta$ be random variables such that
	\begin{equation*}
		x \mid m, s^2, \varphi^2 \sim  \mathcal{N}\left(m, s^2 \varphi^2 \right) \quad \text{and} \quad \varphi^2 \sim \mathcal{IG}\left(\frac{v}{2}, \frac{v}{2}\right),
	\end{equation*}
	then $x \mid m, s^2, v \sim  \mathsf{t}_v\left(m, s^2\right)$ \citep{Geweke-1993}.
	\label{prop2}
\end{prop}
\noindent Following Propositions \ref{prop1} and \ref{prop2}, after the introduction of a latent factor $\varphi_j^2 \sim \mathcal{IG}(1/2, 1/2)$, the prior distribution of $\lambda_j$ can be conveniently re-written as a normal distribution truncated at $0$. This way the normal prior for co-data regression coefficients $\boldsymbol{\gamma}$ is conjugated and the parameters can easily be updated by sampling from a multivariate normal distribution. The set of prior distributions for parameters $\lambda_j$, $\varphi_j^2$, $\boldsymbol{\gamma}_d$ can be re-written as
\begin{align*}
	\lambda_j \mid \mathbf{Z}, \boldsymbol{\gamma}, \varphi_j^2 \sim & \; \mathcal{N} \left(\sum_{d = 1}^D \left(\mathbf{z}_j^d\right)^\intercal \boldsymbol{\gamma}_d, s_0^2 \varphi_j^2 \right) \cdot \mathbb{I}_{\left(\lambda_j > 0\right)}, \\
	\varphi_j^2 \sim & \; \mathcal{IG}\left(\frac{1}{2}, \frac{1}{2}\right), \quad j = 1, \dots, p, \\
	\boldsymbol{\gamma}_d \mid \kappa_d^2 \sim & \; \mathcal{N}_{m_d} \left(\mathbf{0}, \kappa_d^2\mathbf{I}_{m_d}\right), \\
    \kappa_d^2 \sim &\; \mathcal{IG} \left(a_d, b_d\right), \quad d = 1, \dots, D.
\end{align*}
Let $\mu_j = \sum_{d = 1}^D \left(\mathbf{z}_j^d \right)^\intercal \boldsymbol{\gamma}_d$, $\boldsymbol{\lambda} = \left[\lambda_1 \dots \lambda_p\right]^\intercal$ and $\boldsymbol{\Phi}^2 = \text{diag}\left(\boldsymbol{\varphi}\right)$, with $\boldsymbol{\varphi} = \left[\varphi_1^2, \dots, \varphi_p^2\right]^\intercal$. The full-conditional distributions of $\lambda_j$, $\boldsymbol{\gamma}$, $\varphi_j^2$ and $\kappa_d^2$ in the augmented model are
\begin{equation}
\label{eq:postlambda_j}
\begin{aligned}
        \pi\left(\lambda_j \mid \mathbf{Z}, \beta_j, \sigma^2, \tau^2, \boldsymbol{\gamma}, \varphi_j^2\right) \propto & \;\lambda_j^{-1} e^{-\frac{\beta_j^2}{2\sigma^2\tau^2\lambda_j^2} - \frac{\lambda_j^2}{2 s_0^2\varphi_j^2} + \frac{\mu_j \lambda_j}{s_0^2 \varphi_j^2}} \cdot \mathbb{I}_{\left(\lambda_j > 0\right)},  \\
	\varphi_j^2 \mid \mathbf{Z}, \lambda_j, \boldsymbol{\gamma} \sim & \; \mathcal{IG}\left(1, \frac{1}{2} + \frac{(\lambda_j - \mu_j)^2}{2 s_0^2}\right), \\
	\boldsymbol{\gamma} \mid \mathbf{Z}, \boldsymbol{\lambda}, \boldsymbol{\varphi}, \boldsymbol{\kappa}^2 \sim & \; \mathcal{N}_{M} \left(\boldsymbol{\Sigma}_{\gamma}^\star \left(\mathbf{Z}^\intercal \boldsymbol{\Phi}^{-2} \boldsymbol{\lambda}\right), s_0^2 \boldsymbol{\Sigma}_{\gamma}^\star\right), \\
	\kappa_d^2 \mid \boldsymbol{\gamma}_d \sim & \; \mathcal{IG} \left(a_d + \frac{m_d}{2}, b_d + \frac{\boldsymbol{\gamma}_d^\intercal\boldsymbol{\gamma}_d}{2}\right), 
\end{aligned}
\end{equation}
where $\boldsymbol{\Sigma}_{\gamma}^\star = \left( \mathbf{Z}^\intercal \boldsymbol{\Phi}^{-2} \mathbf{Z} + s_0^2\mathbf{D}_{\kappa}^{-1}\right)^{-1}$ and $\mathbf{D}_{\kappa} = \text{diag}\left(\kappa_1^2 \mathbf{1}_{m_1}, \dots, \kappa_D^2 \mathbf{1}_{m_D}\right)$. Details for sampling parameters $\lambda_j$ without computing the unknown normalizing constant are given in Section \ref{sec:rejection} of Supporting Information.

The introduced framework presents a computational advantage: the local variances $\lambda_j^2$ can be computed in parallel, potentially improving the efficiency of the model.

\vspace{0.4cm}
\noindent \textbf{\textit{3. Sampling $\tau^2$.}}
The half-Cauchy prior for the global scale parameter $\tau$ is not conjugated to the prior variance of $\beta_j$, $j = 0, \dots, p$, in a linear regression model with gaussian errors. Therefore, we rely on the strategy proposed by \cite{Makalic-2016} in order to update $\tau$. The authors point out that the half-Cauchy distribution can be written as a scale mixture of inverse-Gamma distributions, which allows conjugate updates of $\tau^2$.

\begin{prop}
	If $\tau^2 \mid \zeta \sim \mathcal{IG}\left(\frac{1}{2}, \frac{1}{\zeta} \right)$ and $\zeta \sim \mathcal{IG}\left(\frac{1}{2}, 1\right)$, then $\tau \sim \mathcal{C}^+(0, 1)$.
	\label{prop4}
\end{prop}

\noindent Following proposition \ref{prop4} the full-conditional distributions for $\tau^2$ and $\zeta$ are available in closed-form. In particular the parameters are updated by sampling from the following densities:
\begin{align*}
	\tau^2 \mid \boldsymbol{\beta}, \sigma^2, \lambda_0^2, \boldsymbol{\lambda}, \zeta \sim & \; \mathcal{IG} \left(\frac{p}{2} + 1, \frac{1}{\zeta} + \frac{\beta_0^2}{2\sigma^2\lambda_0^2} + \frac{1}{2\sigma^2}\sum_{j=1}^p \frac{\beta_j^2}{\lambda_j^2} \right), \\
	\zeta \mid \tau^2 \sim & \; \mathcal{IG}\left(1, 1 + \frac{1}{\tau^2}\right).
\end{align*}

\noindent Section \ref{sec:lambda0} of Supporting Information discusses the update of parameter $\lambda_0^2$, the local scale parameter of the intercept $\beta_0$, using a similar proposition as \ref{prop4}.

\section{Variational Bayes approximation}
\label{sec:VB}

When the number of covariates is huge the method introduced in Section \ref{sec:model} becomes computationally infeasible. In this section an efficient approximation of the joint posterior distribution is discussed. We refer to Section \ref{app:variational} of Supporting Information for the derivation of the Variational approximation.

Under the assumptions of the model introduced in Section \ref{sec:model}, the mean field approximation yields
\begin{align*}
	q\left(\boldsymbol{\theta}\right) =& \; q\left(\boldsymbol{\beta}\right) \cdot q\left(\lambda_0^2\right) \cdot q\left(\lambda_1\right) \cdot \hdots \cdot q\left(\lambda_p\right) \cdot q\left(\psi_0\right) \cdot q\left(\varphi^2_1\right) \cdot \hdots \cdot q\left(\varphi^2_p\right) \cdot \nonumber \\
	& \; \; q\left(\boldsymbol{\gamma}\right) \cdot q\left(\kappa^2_1\right) \cdot \hdots \cdot q\left(\kappa^2_D\right) \cdot q\left(\tau^2\right) \cdot q\left(\zeta\right) \cdot q\left(\sigma^2\right),
\end{align*}
where $q\left(\boldsymbol{\beta}\right)$ and $q\left(\boldsymbol{\gamma}\right)$ denote the joint variational distribution of $\beta_0, \dots, \beta_p$ and $\gamma_1, \dots, \gamma_M$, respectively. At each iteration of the algorithm, the variational factors are updated as
\begin{align*}
	q^\star\left(\boldsymbol{\beta}\right) = &\; \mathcal{N}_{p+1}\left(\boldsymbol{\mu}_\beta^\star, \mathbb{E}_{\sigma^2}\left[\sigma^2\right] \boldsymbol{\Sigma}_\beta^\star\right), \\ & \; \boldsymbol{\mu}_\beta^\star = \boldsymbol{\Sigma}_\beta^\star \mathbf{X}^\intercal \mathbf{y}, \quad \boldsymbol{\Sigma}_\beta^\star = \left(\mathbf{X}^\intercal \mathbf{X} + \mathbb{E}_{\lambda_0^2 \cdot \lambda \cdot \tau^2} \left[\tau^{-2} \boldsymbol{\Lambda}^{-2}\right]\right)^{-1},  \nonumber \\
	q^\star\left(\lambda_0^2\right) = &\; \mathcal{IG} \left(1, a_0^\star\right),  \\ &\; a_0^\star =  \mathbb{E}_\zeta \left[\psi_0^{-1}\right] + \frac{1}{2}\mathbb{E}_{\beta_0 \cdot \sigma^2 \cdot \tau^2}\left[\frac{\beta_0^2}{\sigma^2 \tau^2}\right], \nonumber \\
	q^\star\left(\psi_0\right) = & \; \mathcal{IG}\left(1, k_0^\star\right),  \\ & \; k_0^\star = 1 + \mathbb{E}_{\lambda_0^2}\left[\lambda_0^{-2}\right], \nonumber \\
	q^\star\left(\lambda_j\right) \propto &\; \lambda_j^{-1} \exp\left\{-\frac{a_j^\star}{\lambda_j^2} - b_j^\star \lambda_j^2 + c_j^\star \lambda_j\right\} \cdot \mathbb{I}_{\left(\lambda_j > 0\right)}, \quad j = 1, \dots, p,  \\ 
	& \; a_j^\star = \frac{1}{2} \mathbb{E}_{\beta \cdot \sigma^2 \cdot \tau^2} \left[\frac{\beta_j^2}{\sigma^2 \tau^2}\right], \quad b_j^\star =\frac{1}{2s_0^2} \mathbb{E}_{\varphi^2}\left[\varphi_j^{-2}\right],
	\quad c_j^\star = \frac{1}{s_0^2} \mathbf{z}_j^\intercal \mathbb{E}_{\gamma \cdot \varphi^2} \left[\frac{\boldsymbol{\gamma}}{\varphi_j
	^2}\right],  \nonumber \\
	q^\star\left(\varphi_j^2\right) = & \; \mathcal{IG} \left(1,  d_j^\star \right), \\ & \; d_j^\star = \frac{1}{2} + \frac{1}{2s_0^2} \mathbb{E}_{\lambda \cdot \gamma}\left[\left(\lambda_j - \mathbf{z}_j^\intercal \boldsymbol{\gamma}\right)^2\right], \nonumber \\
	 q^\star\left(\boldsymbol{\gamma} \right) = & \;  \mathcal{N}_M \left(\boldsymbol{\mu}_\gamma^\star, s_0^2\boldsymbol{\Sigma}_\gamma^\star\right), \\ & \; \boldsymbol{\mu}_\gamma^\star = \boldsymbol{\Sigma}_\gamma^\star \mathbf{Z}^\intercal \mathbb{E}_{\varphi^2 \cdot \gamma}\left[\boldsymbol{\Phi}^{-2} \boldsymbol{\lambda}\right], \quad \boldsymbol{\Sigma}_\gamma^\star = \left( \mathbf{Z}^\intercal \mathbb{E}_{\varphi^2}\left[\boldsymbol{\Phi}^{-2}\right] \mathbf{Z} + s_0^2 \mathbb{E}_{\kappa^2}\left[\mathbf{D}_\kappa^{-1}\right]\right)^{-1}, \nonumber \\
	q^\star\left(\kappa_d^2\right) = &\; \mathcal{IG} \left( e_d^\star, f_d^\star\right), \quad d = 1, \dots, D, \\ & \; e_d^\star = a_d + \frac{m_d}{2},\quad  f_d^\star = b_d + \frac{1}{2}  \mathbb{E}_\gamma \left[\boldsymbol{\gamma}_d^\intercal \boldsymbol{\gamma}_d\right] ,  \nonumber \\
	q^\star\left(\tau^2\right) = &\; \mathcal{IG} \left(\frac{p}{2} + 1, g^\star\right),  \\ &\; g^\star =  \mathbb{E}_\zeta \left[\zeta^{-1}\right] + \frac{1}{2}\mathbb{E}_{\beta_0 \cdot \lambda_0^2 \cdot \tau^2}\left[\frac{\beta_0^2}{\tau^2 \lambda_0^2}\right] + \frac{1}{2}  \sum_{j=1}^p  \mathbb{E}_{\beta \cdot \sigma^2 \cdot \lambda} \left[\frac{\beta_j^2}{\sigma^2 \lambda_j^2}\right], \nonumber
\end{align*}

\newpage

\begin{align*}
	q^\star\left(\zeta\right) = & \; \mathcal{IG}\left(1, h^\star\right),  \\ & \; h^\star = 1 + \mathbb{E}_{\tau^2}\left[\tau^{-2}\right], \nonumber \\
	q^\star\left(\sigma^2\right) = & \; \mathcal{IG}\left(v + \frac{n+p+1}{2}, l^\star\right),  \\ & \; l^\star = q + \frac{1}{2} \left( \mathbb{E}_{\beta}\left[\Vert\mathbf{y} - \mathbf{X}\boldsymbol{\beta}\Vert_2^2\right] + \mathbb{E}_{\beta_0 \cdot \lambda_0^2 \cdot \tau^2}\left[\frac{\beta_0^2}{\tau^2 \lambda_0^2}\right] + \sum_{j = 1}^p \mathbb{E}_{\beta \cdot \lambda \cdot \tau^2}\left[\frac{\beta_j^2}{\tau^2 \lambda_j^2}\right]\right), \nonumber
\end{align*}
where the expectations are taken with respect to the updated variational family $q^\star\left(\boldsymbol{\theta}\right)$. The variational lower bound $\mathcal{L}$ of the marginal distribution $p\left(\mathbf{y}\right)$ is computed as
\begin{align}
	\mathcal{L}	\propto &\; \frac{1}{2}\left(\log \vert\boldsymbol{\Sigma}_\beta^\star\vert + \log \vert\boldsymbol{\Sigma}_\gamma^\star\vert\right) + \frac{p+1}{2} \mathbb{E}_{\sigma^2}\left[\log\sigma^2\right] - \left(v + \frac{n+p+1}{2} \right) \log l^\star +  \nonumber \\
	& \; \sum_{j=1}^p \left(\log s_j - \log k_j + a_j^\star \mathbb{E}_{\lambda}\left[\lambda_j^{-2}\right] + b_j^\star \mathbb{E}_{\lambda}\left[\lambda_j^2\right] - c_j^\star \mathbb{E}_{\lambda}\left[\lambda_j\right] - \log d_j^\star \right) -  \nonumber \\
	& \;  \sum_{d=1}^D e_d^\star \log f_d^\star - \left(\frac{p}{2} + 1\right) \log g^\star - \log h^\star - \log a_0^\star - \log k_0^\star,
	\label{eq:lb}
\end{align}
where $k_j = 1 - \mathbb{P}\left(\mathcal{N}\left(\mathbf{z}_j^\intercal \boldsymbol{\gamma}, s_0^2 \varphi_j^2\right) < 0 \right)$ and $s_j$ is the unknown normalizing constant of parameters $\lambda_j$ in \eqref{eq:postlambda_j}. Each component is derived in Section \ref{app:a1} of Supporting Information.

As opposed to the Gibbs sampler, the Variational algorithm does not require the explicit inversion of the $p \times p$ matrix $\boldsymbol{\Sigma}_\beta^\star$, as only quantities $\boldsymbol{\Sigma}_\beta^\star \mathbf{X}^\intercal\mathbf{y}$, $\text{diag}\left(\boldsymbol{\Sigma}_\beta^\star\right)$ and $\mathbf{X} \boldsymbol{\Sigma}_\beta^\star \mathbf{X}^\intercal$ are needed. These can be efficiently evaluated with complexity $\mathcal{O}\left(n^2 p\right)$ following the work of \cite{Munch-2019}. The algebraic details are shown in Section \ref{app:b1} of Supporting Information. The bottleneck of the algorithm is the computation of the terms $\log s_j$, $\mathbb{E}_\lambda \left[\lambda_j\right]$, $\mathbb{E}_\lambda \left[\lambda_j^2\right]$ and $\mathbb{E}_\lambda \left[\lambda_j^{-2}\right]$, since no closed-form is available. These can be evaluated in parallel following the procedure described in Section \ref{app:variational} of Supporting Informatiion.

\section{Simulation study}
\label{sec:simulation}
%

In this section we empirically show the quality of the Variational approximation to the joint posterior distribution and the benefit of using co-data for variable selection. We use a simulation scheme that resembles the one by \cite*{Johnson-2013}. In particular, we assess the effectiveness of the approximation on low to moderate $p$ problems, whereas variable selection is evaluated on higher dimensional settings. We refer to Section \ref{sec:sim_scheme} of Supporting Information for further details on the data generation scheme.

Variable selection is evaluated with the average area under the ROC curve ($AUC$) of the selected variable, where the posterior inclusion probability of the generic covariate $j$ is evaluated with the quantity $\mathbb{E}_{\lambda_j} \left[\lambda_j^2 / \left(1 + \lambda_j^2\right)\right]$ \citep{Carvalho-2010}.

The hyperparameters for $\sigma^2$ and $\kappa_1^2$ are set to $(v, q) = (1, 10)$ and $(a_1, b_1) = (1, 10)$, respectively. The Gibbs sampler is run for $B = 5000$ iterations with a burn-in period of $bn = 2500$ iterations, while the Variational algorithm is stopped either if the increase of the lower bound is less than $\epsilon = 0.001$ or if the maximum number of iteration $B = 1000$ is reached.

\vspace{0.4cm}
\noindent \textbf{\textit{1. Performance of variable selection.}}

\noindent Here we analyse how the model behave as the degree of prior information varies. We study the cases $n = (50, 100, 150)$ and $p = (500, 1000, 1500)$ and we set the number of true non-zero coefficients to $p_0 = 30$. To study how the model borrows information from the co-data, we simulate co-data sources representing five degrees of prior information:
\begin{itemize}
    \item[] $G0\big)$ \textbf{no co-data} set-up: we include in the co-data matrix only the intercept, therefore $\mathbf{Z} = \mathbf{1}_p$;
    \item[] $G1\big)$ \textbf{non-informative} set-up: a binary co-data source is included in the model by randomly selecting $100$ regressors, therefore the co-data matrix $\mathbf{Z}$ is created from the binary vector $\mathbf{z} \in \left\{0, 1\right\}^p$, where $z_j = 1$ if the $j$-th variable is selected, $z_j = 0$ otherwise;
    \item[] $G2\big)$ \textbf{weakly informative} set-up: a binary co-data source is included in the model by randomly selecting $20$ of the true non-zero regressors and $80$ of the true zero regressors, therefore the co-data matrix $\mathbf{Z}$ is created from the binary vector $\mathbf{z} \in \left\{0, 1\right\}^p$, where $z_j = 1$ if the $j$-th variable is selected, $z_j = 0$ otherwise;
    \item[] $G3\big)$ \textbf{informative} set-up: a binary co-data source is included in the model by randomly selecting $20$ of the true non-zero regressors and $10$ of the true zero regressors, therefore the co-data matrix $\mathbf{Z}$ is created from the binary vector $\mathbf{z} \in \left\{0, 1\right\}^p$, where $z_j = 1$ if the $j$-th variable is selected, $z_j = 0$ otherwise;
    \item[] $G4\big)$ \textbf{perfect co-data information} set-up: the co-data matrix $\mathbf{Z}$ is created from the binary vector $\mathbf{z} \in \left\{0, 1\right\}^p$, where $z_j = 1$ if $\beta_j^0 \ne 0$, $z_j = 0$ otherwise.
\end{itemize}

\begin{figure}
	\centering
	\includegraphics[scale = 0.45]{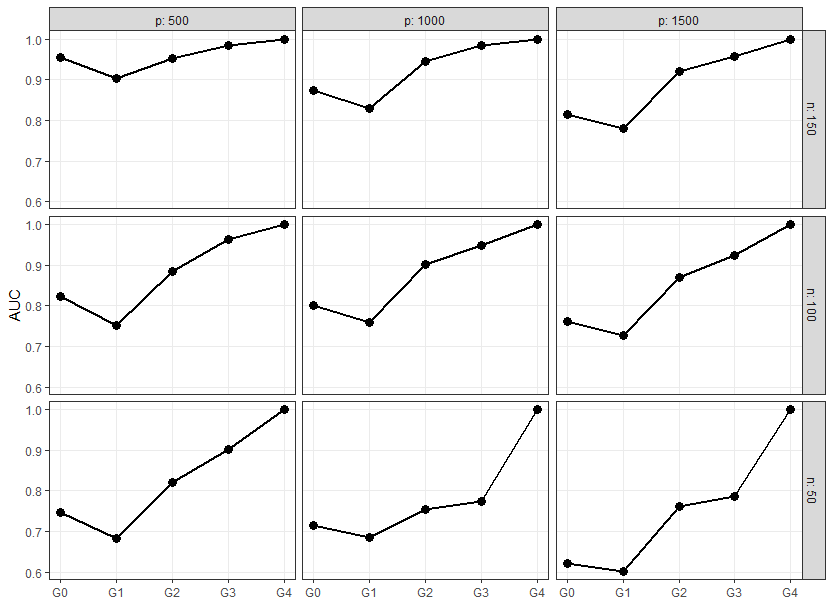}
	\caption{Variable selection with Variational algorithm for co-data scenarios $G0$ to $G4$; $AUC$ is averaged over $25$ replicates of the experiments.}
	\label{fig:sim}
\end{figure}

\noindent The results are shown in Figure \ref{fig:sim}. The model is able to learn from the auxiliary information and the variable selection performance improves when the co-data is actually informative. The $AUC$, indeed, increases alongside the magnitude of prior information and the variables are perfectly selected when the co-data provides perfect information (case $G4$). On the other hand, random co-data is associated with a small loss of performance and leads to lower scores of $AUC$. In particular, if we compare the case with random co-data ($G1$) and no co-data ($G0$), the latter performs slightly better. This difference, however, vanishes as the number of covariates increases. Finally, the smallest increase in $AUC$, as expected, is between $G2$ and $G3$, since these cases represent the most similar degree of co-data information. 

\vspace{0.4cm}
\noindent \textbf{\textit{2. Gibbs sampler vs Variational inference.}}
\label{subsec:GSvsVI}

\noindent The accuracy of the Variational approximation is evaluated by comparing it to the Gibbs sampler in terms of variable selection and mean squared error ($MSE$) between $\boldsymbol{\beta}$ estimates. We analysed the cases $n = (50, 100)$ and $p = (75, 125, 200)$ and we set $p_0 = 30$. We consider four degrees of prior information and we refer to Section \ref{app:c1} of Supporting Information for further details on co-data simulation.

The results of variable selection are shown in Figure \ref{fig:sim1} of Supporting Information. The two methods behave similarly in all the considered cases. In particular, the results are equal for the lowest dimensional case, whereas the Variational approximation approaches the performance of the Gibbs sampler when $p$ increases. Table \ref{tab:mse} reports the details of the similarity between the two methods in terms of $\boldsymbol{\beta}$ estimate. Gibbs sampling and Variational algorithm provide similar estimates for the regression parameters, as the mean of the $MSE$ is close to zero for all the considered cases. In particular, when $n$ is large and the co-data are strongly informative the two methods provide almost the same estimate.

In Figure \ref{fig:sim_sd} of Supporting Information we show the average estimated standard deviations of regression parameters $\boldsymbol{\beta}$ across the different cases of co-data information. The results follows the considerations made in \citet*{MacKay-2003, Wang-2005, Giordano-2017}, as the Variational approximation underestimates the posterior variances when compared to Gibbs sampling. However, as highlighted in \cite{Blei-2006} and from the simulation results, this lack of accuracy does not necessarily affect the global performance of the model. 

The Gibbs sampler presents one main drawback: for some settings of $n$ and $p$, the sampling method presented in Section \ref{sec:rejection} of Supporting Information suffers from low acceptance probability. Therefore, the efficiency of the Gibbs sampler is heavily affected in a negative way. Given the good results of the Variational approximation and the poor performance in terms of computational efficiency of the Gibbs sampler, we rely on the former for the following large $p$ applications.

\begin{table}
\caption{Mean of $MSE_{\boldsymbol{\beta}}$ between Gibbs sampling and the Variational algorithm; the mean is evaluated over $10$ replicates of each experiment.}
\centering
\begin{tabular}{lllll}
\hline
\multicolumn{5}{c}{$MSE_{\boldsymbol{\beta}} = \Vert \boldsymbol{\beta}_{\text{FB}} - \boldsymbol{\beta}_{\text{GS}}\Vert_2^2$}                         \\ \hline \hline
\multicolumn{1}{l}{}     & \multicolumn{1}{l}{$p = 75$} & \multicolumn{1}{l}{$p = 125$} & \multicolumn{1}{l}{$p = 200$} &                            \\ \hline
\multicolumn{1}{l}{$G0$} & \multicolumn{1}{l}{0.0079}   & \multicolumn{1}{l}{0.2592}    & \multicolumn{1}{l}{0.0365}    & \multirow{4}{*}{$n = 50$}  \\ \cline{1-4}
\multicolumn{1}{l}{$G1$} & \multicolumn{1}{l}{0.0074}   & \multicolumn{1}{l}{0.1034}    & \multicolumn{1}{l}{0.0070}    &                            \\ \cline{1-4}
\multicolumn{1}{l}{$G2$} & \multicolumn{1}{l}{0.0091}   & \multicolumn{1}{l}{0.0647}    & \multicolumn{1}{l}{0.0621}    &                            \\ \cline{1-4}
\multicolumn{1}{l}{$G3$} & \multicolumn{1}{l}{0.0032}   & \multicolumn{1}{l}{0.1153}    & \multicolumn{1}{l}{0.0787}    &                            \\ \hline \hline
\multicolumn{1}{l}{$G0$} & \multicolumn{1}{l}{0.0006}   & \multicolumn{1}{l}{0.0268}    & \multicolumn{1}{l}{0.0217}    & \multirow{4}{*}{$n = 100$} \\ \cline{1-4}
\multicolumn{1}{l}{$G1$} & \multicolumn{1}{l}{0.0007}   & \multicolumn{1}{l}{0.0023}    & \multicolumn{1}{l}{0.0212}    &                            \\ \cline{1-4}
\multicolumn{1}{l}{$G2$} & \multicolumn{1}{l}{0.0005}   & \multicolumn{1}{l}{0.0046}    & \multicolumn{1}{l}{0.0023}    &                            \\ \cline{1-4}
\multicolumn{1}{l}{$G3$} & \multicolumn{1}{l}{$< 0.0001$}   & \multicolumn{1}{l}{0.0037}    & \multicolumn{1}{l}{0.0001}    &                            \\ \hline
\end{tabular}
\label{tab:mse}
\end{table}

\section{Application to real data}
\label{sec:realdata}

In this section we present two genomics applications. The first allows to evaluate our method for binary co-data in a multiple linear regression context. The second involves multiple co-data sources in a classification setting with a large number of variables.

In Section \ref{sec:simulation} we show that the thresholding variable selection works well when the covariates are sampled independently. However, the posterior probabilities are treated separately and the optimal threshold is (almost) never $0.5$. For this reason in this section we also consider the more recent selection procedure called \textit{decoupling shrinkage and selection} (DSS) proposed by \cite{Hahn-2015}. This method was introduced to deal with potentially very strong correlations, as they are present in many genomics datasets. The authors propose a posterior variable selection summary based on the posterior mean of the predictors which results in a sequence of sparse models. Shrinkage can be achieved with any prior distribution and is `decoupled' from the selection approach based on the posterior distribution. DSS method relies on the optimization of a loss function which balances the prediction error and the sparseness of the solution. Given an estimate $\hat{\boldsymbol{\beta}}$, the solution is obtained via adaptive LASSO by solving the following optimization problem, 
\begin{equation*}
	\boldsymbol{\theta}^{DSS} = \text{argmin}_{\boldsymbol{\theta}} \; \frac{1}{n} \Vert \mathbf{X}\hat{\boldsymbol{\beta}} - \mathbf{X} \boldsymbol{\theta}\Vert_2^2 + \lambda\sum_{j = 0}^p \frac{\vert\theta_j\vert}{\vert\hat{\beta}_j\vert},
\end{equation*}
where the smoothing parameter $\lambda$ operates as a thresholding parameter and can be estimated with cross-validation over a set of values (grid search). The authors advocate this method over thresholding mainly because it naturally handles multi-collinearity.

\subsection{Case study 1: p38MAPK pathway}

Following \cite{Kpogbezan-2019}, the model is tested on the p38MAPK pathway dataset. We investigate the effect of single nucleotide polymorphisms (SNPs) on the genes in the pathway. A subset of the data collected in the GEUVADIS project \citep{Lappalainen-2013} is considered. For each of the $99$ genes we collect a different number $p_t$ of SNPs, $ t = 1, \dots, 99$, with minimum $p_t = 56$ and maximum $p_t = 1169$. After a first sub-selection, $n = 373$ RNA-Seq samples are obtained from the 1000 Genomes Project. Since it is believed that SNPs within the gene's range have stronger influence on the gene's expression, a binary co-data source $\mathbf{z}_t \in \left\{0, 1\right\}^{p_t}$ is included in the analysis, where $z_{t, j} = 1$ if the $j$-th SNP related to gene $t$ is located inside the gene's range and $z_{t, j} = 0$ otherwise. 

The hyperparameters for $\sigma^2$ and $\kappa_1^2$ are set to $(v, q) = (1, 10)$ and $(a_1, b_1) = (1, 10)$. The algorithm is stopped either if the increase of the variational lower bound is less than $\epsilon = 0.001$ or if the maximum number of iteration $B = 1000$ is reached.

\begin{figure}
	\centering
	\includegraphics[scale = 0.5]{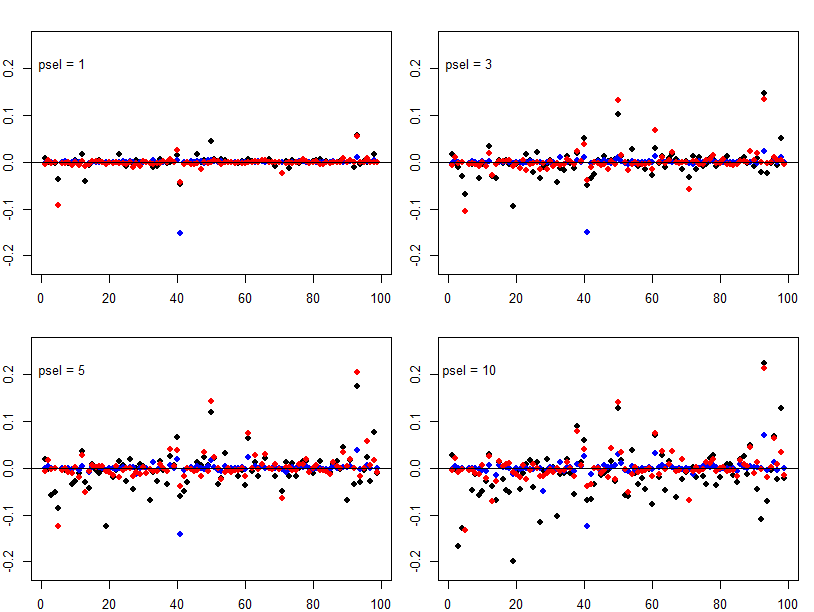}
	\caption{Relative reduction of MSE (rrMSE) of LASSO (darkgrey dots), infHS regression with DSS selection procedure (black dots) and infHS regression with thresholding selection procedure (lightgrey dots) for all the $99$ genes. For each panel, the maximum number of selected SNPs is fixed to $1$, $3$, $5$, and $10$.}
	\label{fig:rmse}
\end{figure}

To evaluate the prediction performance of the model, the data are divided in training-set ($n_1 = 249$) and test-set ($n_2 = 124$). We rely on the relative reduction of the mean squared error (rrMSE) for all the $99$ genes. The rrMSE for the $t$-th regression model is defined as
\begin{equation*}
    \text{rrMSE}_t = 1 - \frac{\text{MSE}_t}{\text{MSE}_0},
\end{equation*}
where $\text{MSE}_0$ and $\text{MSE}_t$ are the mean squared error of the null model and the mean squared error of the $t$-th linear model, respectively. The mean of $\mathbf{y}$ is centered around zero, therefore we do not include the intercept in the null model. Note that larger values of $\text{rrMSE}_t$ are associated with better predictive performance. On the contrary, negative values denote worse predictive performance than the null model. 

We test our method against LASSO regression \citep[]{Tibshirani-1996}. The results are shown is Figure \ref{fig:rmse} for different degrees of sparseness. For the infHS regression, the SNPs are selected with both the thresholding procedure used in Section \ref{sec:simulation} (lightgrey dots) and the DSS method (black dots). The latter provides a better prediction for all genes when compared with the thresholding variable selection. This is justified by the strong correlations among the SNPs. When comparing infHS and LASSO regressions, instead, the considerations made in \cite{Kpogbezan-2019} hold. The results of the LASSO are more noisy, likely due to less shrinkage of the near-zero coefficients, whereas most of the $\text{rrMSE}_t$ are concentrated around zero for the infHS model. The effect of the SNPs are relevant only for a small number of genes: the largest values of $\text{rrMSE}_t$ are associated to genes $50$, $61$ and $93$. Both infHS and LASSO are able to capture these effects, however the latter is more prone to negative values of $\text{rrMSE}_t$. The LASSO gives better results for gene $98$. Note that the method by \cite{Kpogbezan-2019} gives fairly similar results to ours, as it is also based on the Horseshoe. That method, however, is much more limited in use, as it only handles one discrete co-data source and only continuous outcomes. 

Our method take $68$ minutes to estimate all the $99$ regressions and to evaluate the different variable selection approaches. The algorithm is run on a x64 Windows 11 operating system.

\subsection{Case study 2: methylation data}

The model is tested on the methylation dataset of \citet{Verlaat-2018}, which contains methylation profiles of self-collected cervicovaginal lavages corresponding to $28$ women with normal cervix and $36$ women with high-grade precursor lesions (CIN3), for a total of $n = 64$ samples. In order to improve the diagnostic classification we include in the analysis $D = 5$ co-data sources: the standard deviations of the probes, p-values from the previous study, a binary variable differentiating the hypo-methylated and hyper-methylated probes, a categorical variable with $6$ categories denoting the genomic region of each probe and the probes' means in cancer cells. See Section \ref{app:probit} of Supporting Information for theorical details of the Variational approximation to the joint posterior distribution in a probit regression model. Based on the previous results, we consider the probes with an external p-value $p_\alpha < 0.005$, resulting in a total of $p = 11251$ probes, where each probe measures the methylation of a unique location on the genome. 

\begin{figure}
    \centering
    \subfloat{{\includegraphics[width=.36\textwidth]{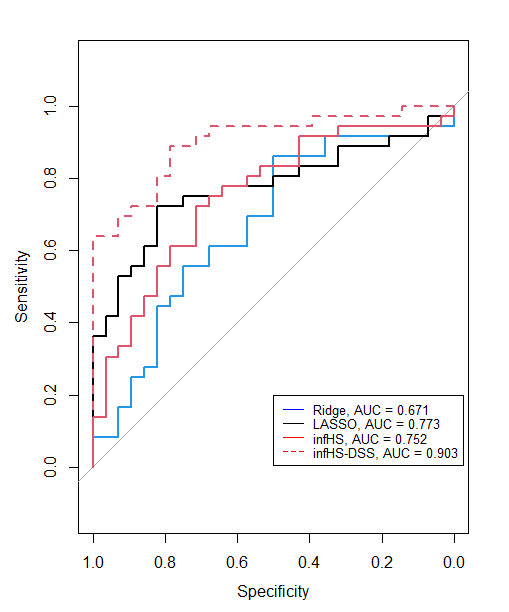} }}
    \quad
    \subfloat{{\includegraphics[width=.59\textwidth]{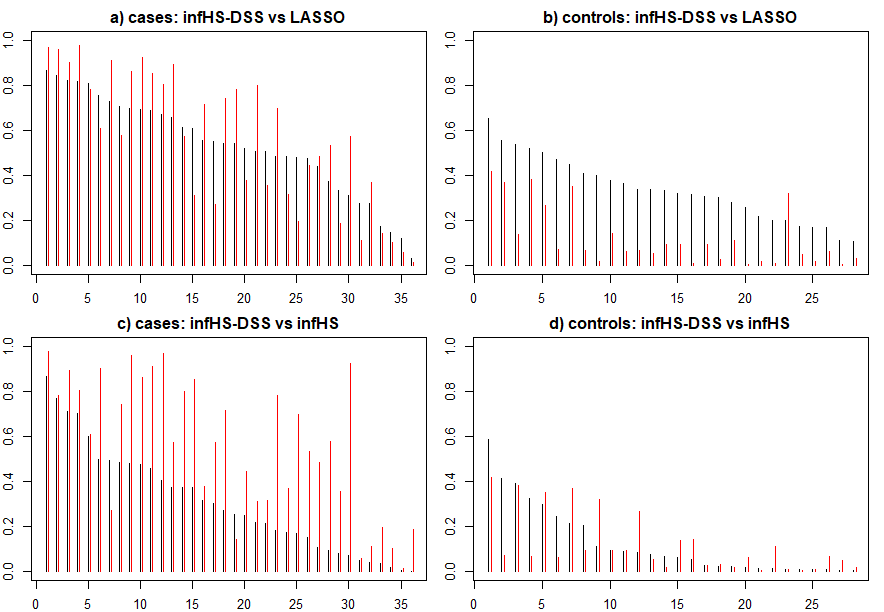} }}
    \caption{Results of the LOOCV. Left panel: ROC curves for Ridge regression (lightgrey), LASSO (darkgrey), the informative Horseshoe regression (black) and the informative Horseshoe with DSS variable selection procedure (dotted-black); right panel: \textit{a} and \textit{b.} predicted probabilities for cases ($y_i = 1$) and controls ($y_i = 0$) with infHS-DSS (black) and LASSO (darkgrey) in decreasing order of LASSO prediction; \textit{c} and \textit{d.} predicted probabilities for cases and controls with infHS-DSS (black) and ordinary infHS (darkgrey) in decreasing order of infHS prediction.}
    \label{fig:roc}
\end{figure}

We test our model against the ordinary Ridge regression and the LASSO.  The hyperparameters are set to $\left(a_d, b_d\right) = (1, 10)$ for $d = 1, \dots, 5$ and the predictive performances are evaluated by leave-one-out cross-validation (LOOCV). As for case study 1, the algorithm is stopped either if the increase of the variational lower bound is less than $\epsilon = 0.001$ or if the maximum number of iteration $B = 1000$ is reached. We include in the analysis also the thresholding and the DSS versions of infHS model. The former does not exclude any of the covariates from the model, since all the posterior inclusion probabilities are greater than 0.5. Therefore the results of this approach are not shown. We also fit our model without co-data; due to many cases ($y_i=1$) corresponding to very low predicted probabilities, the results are not competitive to those of infHS, and hence not shown here. 

The left panel of Figure \ref{fig:roc} shows the ROC curves for the considered competitors, where the predicted probabilities for infHS are computed as $p_i = \Phi\left(\mathbf{x}_i^\intercal \boldsymbol{\beta}\right)$. InfHS-DSS gives the best forecasting results with $\text{AUC} = 0.903$, greatly improving compared to its ordinary version with $\text{AUC} = 0.752$; the LASSO performs better than infHS with $\text{AUC} = 0.773$ and provides the sparsest model, with an average of $8$ variables included as opposed to the $49$ estimated by infHS-DSS. Among these $49$ probes, $37$ of them are selected in more than $70\%$ of the folds, whereas only $10$ appear in all. The ordinary Ridge does not compete in terms of predictive performance ($\text{AUC} = 0.671$). Note that GRridge co-data method \citep{vandeWiel-2016}, which is one of the few methods that also tackles multiple co-data sources, was previously reported to achieve an $\text{AUC} = 0.77$ \citep[]{Verlaat-2018}. 

The right panel of Figure \ref{fig:roc} shows the predicted probabilities of infHS-DSS model against the LASSO and the ordinary infHS. When compared to the LASSO (figures \textit{a} and \textit{b}, values in decreasing order of LASSO prediction), infHS-DSS gives similar predictions for the cases ($y_i = 1$), with some of the observations achieving higher predictive scores and other assuming lower ones; on the other hand, it provides significantly lower scores for the controls ($y_i = 0$), reducing the false positive rate and improving the overall performance. The impact of the DSS procedure on the prediction can be assessed by analyzing figures \textit{c} and \textit{d}, which compares infHS-DSS to infHS model (values in decreasing order of infHS prediction). The former completely overwhelms the latter, as it gives higher probabilities for almost all cases and reduces the predicted probabilities for the highest scores of the controls. To sum up these results we compute the mean absolute difference between the true labels and the predicted probabilities. The results are shown in Table \ref{tab:pi}, where infHS-DSS provides the best performance.

\begin{table}
\caption{Mean absolute difference between true labels ($y_i = 0$ or $y_i = 1$) and the predicted probabilities ($p_i = \Phi\left(\mathbf{x}_i^\intercal \boldsymbol{\beta}\right)$).}
    \centering
    \begin{tabular}{lllll}
        \hline
                                       & Ridge  & LASSO  & infHS    & infHS-DSS \\ \hline   \hline
        $\sum_{i = 1}^{64} \frac{\vert y_i - p_i\vert}{64}$ & 0.466 & 0.410 & 0.444 & \textbf{0.299}  \\ \hline
    \end{tabular}
    \label{tab:pi}
\end{table}

We summarize the co-data information in Table \ref{tab:gamma}, which shows the distribution of the overall co-data sources and the distribution of the $p_{sel} = 37$ most selected probes. The co-data related to the the probes’ means in cancer cells show the strongest difference in distribution between the original and the selected features, suggesting that this source is the most informative in terms of prior information. No evident differences are found in the other co-data sources.

The computational time is around 5 minutes for the estimation of infHS. The method is tested on a x64 Windows 11 operating system and the local variances are evaluated in parallel with 4 cores.

\vspace{0.4cm}
\begin{table}
\caption{Summary of the co-data estimates and the co-data distribution in the $p_{sel} = 37$ most selected probes and in the original population.}
\centering
    \begin{tabular}{lll}
    \hline
    \multirow{ 2}{*}{Co-data} & Selected probes distribution & Original distribution \\ 
    & ($p_{sel} = 37$) & ($p = 11251$) \\ [0.5ex] \hline \hline
    \multirow{ 3}{*}{external $p$-value} & mean: $0.0016$ & mean: $0.0019$ \\
    & sd: $0.0014$ & sd: $0.0015$ \\ 
    & range: ($2.4\cdot 10^{-6} , 0.005$) & range: ($9.9 \cdot 10^{-11} , 0.005$) \\ \hline
    \multirow{ 3}{*}{probes sd} & mean: $0.035$ & mean: $0.038$ \\
    & sd: $0.022$ & sd: $0.019$ \\ 
    & range: ($0.014 , 0.129$) & range: ($0.009 , 0.320$) \\ \hline
    \multirow{ 6}{*}{genomic region} & Distant: $35.1\%$ & Distant: $31.7 \%$ \\ 
    & Island: $43.2 \%$ & Island: $37.3 \%$ \\
    & N shelf: $5.4 \%$ & N shelf: $4.6 \%$ \\
    & N shore $10.8 \%$ & N shore: $12.4 \%$ \\
    & S shelf: $0 \%$ & S shelf: $4 \%$ \\
    & S shore: $5.4 \%$ & S shore: $10 \%$ \\ \hline
    \multirow{ 2}{*}{degree of methylation} & Hypo-methylated: $ 57 \%$ & Hypo-methylated: $56 \%$ \\
    & Hyper-methylated: $43 \%$ & Hyper-methylated: $44 \%$ \\ \hline
    \multirow{ 3}{*}{cancer cells mean} & mean: $0.150$ & mean: $-0.099$ \\
    & sd: $1.116$ & sd: $1.137$ \\
    & range: ($-1.28 , 3.34$) & range: ($-4.89 , 8.10$) \\ \hline
    \end{tabular}
\label{tab:gamma}
\end{table}

%

\section{Discussion}
\label{sec:conclusions}
%

We introduced a novel Bayesian regression approach  for high-dimensional data able to learn from auxiliary prior information, i.e. co-data. We showed that both prediction and variable selection benefit from the inclusion of co-data, when these are actually informative. The model allows for both continuous and binary outcome, as well as both continuous and discrete co-data sources. In particular, we provided a flexible method that estimates multiple co-data coefficients jointly, contrary to the previous method of \cite{vanNee-2021} that models each source separately.

We discussed a full Bayesian approach, for which we developed a Gibbs sampler to update each parameter iteratively by sampling from their full-conditional distributions. To do so, we introduced a novel rejection sampling method to sample the local variances, which showed a non-closed form target density. Eventually, we proposed a Variational approximation to the joint posterior distribution and applied the CAVI algorithm to optimize the target density. This latter method is suited for high-dimensional problems, as it does not require the explicit inversion of the $p \times p$ posterior covariance matrix $\boldsymbol{\Sigma}_\beta^\star$. In particular, only its diagonal is required and we implemented the methods proposed in \cite{Munch-2019} to efficiently achieve this. Beside this, another computational advantage is the parallel evaluation of the local variances. The evaluation of these parameters represents the computational bottleneck of the method: the acceptance probability of the rejection sampling is small for some settings of the parameters, whereas the numerical integration required in the Variational approximation is the most expensive step of the algorithm.

The Variational inference is less useful than the Gibbs sampler for posterior inference, as it provides posterior means point estimates and underestimates the posterior variances \citep{MacKay-2003, Wang-2005, Turner-2011, Giordano-2017}. This lack of accuracy, on the other hand, does not necessarily affect the predictive performance of the model \citep{Blei-2006}. Therefore, the Variational algorithm should be used for large $p$ datasets, whereas the Gibbs sampling could be useful when the interest is in the posterior inference (credible intervals) and the number of covariates is moderate. 

A possible limitation of infHS for some applications occurs when a categorical co-data source contains one very strong, relatively small co-data group. Under this circumstance, the sparsity assumption may not be realistic for this particular group of variables. For such applications, an interesting extension of infHS would be to allow a dense prior for a small group of variables, for which one expects a particularly relevant prior evidence.

Another extension to allow more flexibility would be the specification of the prior $\lambda_j \sim \text{Half-}t(v)$, with $v > 1$. Here $v = 1$ since the purpose is to apply the Horseshoe prior. In \citet{Biswas-2021} the authors develop coupling techniques for high-dimensional regression with this particular prior and argue that larger values of $v$ can affect the statistical and computational performance of the model. 

To conclude, we provided one of the fastest implementation of Horseshoe regression able to learn from auxiliary information. The R code for \texttt{infHS} and the datasets used are available at \href{https://github.com/cbusatto/infHS}{https://github.com/cbusatto/infHS}.

\section*{Acknowledgements}

The first author was supported also by the project THE - Tuscany Health Ecosystem. The authors thank professor Francesco C. Stingo for his help with the start of this project. \vspace*{-8pt}


\bibliographystyle{apalike}
\bibliography{infHS_biblio}

\appendix

\section{Web Appendix A}
\label{sec:A}

\subsection{Gibbs step for local variance $\lambda_0^2$}
\label{sec:lambda0}
%

The half-Cauchy prior for the local shrinkage parameter $\lambda_0$ of the intercept $\beta_0$ is not conjugated to the variance in a linear regression model with normal errors. We rely on the data-augmentation step proposed in \cite{Makalic-2016} in order to easily and efficiently update parameter $\lambda_0$. 

\begin{prop}
	If $\lambda_0^2 \mid \psi_0 \sim \mathcal{IG}\left(\frac{1}{2}, \frac{1}{\psi_0} \right)$ and $\psi_0 \sim \mathcal{IG}\left(\frac{1}{2}, 1\right)$, then $\lambda_0 \sim \mathcal{C}^+(0, 1)$ \citep{Makalic-2016}.
	\label{prop3}
\end{prop}

\noindent The inverse-Gamma distribution is conjugated to itself and to the local scale parameter, therefore a closed-form full-conditional is available and a Gibbs step can be implemented. The full-conditional distributions of $\lambda_0^2$ and $\psi_0$ are
\begin{align*}
    \lambda_0^2 \mid \beta_0, \tau^2, \sigma^2 \sim & \; \mathcal{IG}\left(1, \frac{1}{\psi_0} + \frac{\beta_0^2}{2\sigma^2 \tau^2}\right), \\
    \psi_0 \mid \lambda_0^2 \sim & \; \mathcal{IG}\left(1, 1 + \frac{1}{\lambda_0^2} \right).
\end{align*}

\section{Web Appendix B}
\label{sec:B}
\subsection{Rejection sampling for parameters $\lambda_j$}
\label{sec:rejection}
%

In this section we propose a novel rejection sampling algorithm to sample shrinkage parameters $\lambda_j$, $j = 1, \dots, p$, from the full-conditional distribution without knowing the normalizing constant. The rejection sampling allows to draw a new value $x_\star$ from a density $f$ by sampling it from a proposal distribution $g$ and accepting it with a probability proportional to the ratio $r(x_\star) = f(x_\star) / g(x_\star)$. In particular, $g$ must be chosen such that the support of $f$ is a subset of the support of $g$. Let
\begin{equation*}
	k = \text{sup}_x \frac{f(x)}{g(x)} < \infty
\end{equation*}
and accept $x_\star$ with probability $a = f(x_\star) / (k g(x_\star))$. It can be shown that the acceptance probability of the algorithm is $1/k$. Therefore, the goal is to find a proposal distribution $g$ such that $k$ is small. The rejection sampling works also if the normalizing constant of $f$ is unknown, as long as a sampling method for $g$ is available. 

Consider the following density
\begin{equation*}
	f(x) = c_f x^{-1} e^{-\psi / x^2 - \alpha^2 x^2 + \beta x} \cdot \mathbb{I}_{(x > 0)},
\end{equation*}
where $\psi, \alpha^2 > 0$, $\beta \in \mathbb{R}$ and $c_f$ is the unknown normalizing constant. New values are sampled from the proposal distribution $g(x) \sim \mathcal{G}_{3p} \left(\gamma, \alpha, \beta\right)$, with density $g(x) = c_g x^\gamma e^{-\alpha^2 x^2 + \beta x}$, where $\gamma \in \mathbb{N}^+$. We set parameters $\alpha$ and $\beta$ equal in both $f$ and $g$ in order for the ratio $r(x)$ to be analytically tractable and easily maximized. Eventually, the acceptance probability is optimized through the choice of parameter $\gamma$. This choice of $\alpha$ and $\beta$ yields
\begin{equation*}
	r(x) = \frac{c_f}{c_g} x^{-\gamma-1} e^{-\psi/x^2} \cdot \mathbb{I}_{(x > 0)},
\end{equation*}
which has one positive maximum at $\dot{x} = \sqrt{2\psi / (\gamma+1)}$. Theoretical properties and details for sampling from a $\mathcal{G}_{3p}$ distribution are shown in \cite{Busatto-2023}. 

\begin{figure}
	\centering
	\vspace{-1.3cm}
	\includegraphics[scale = 0.6]{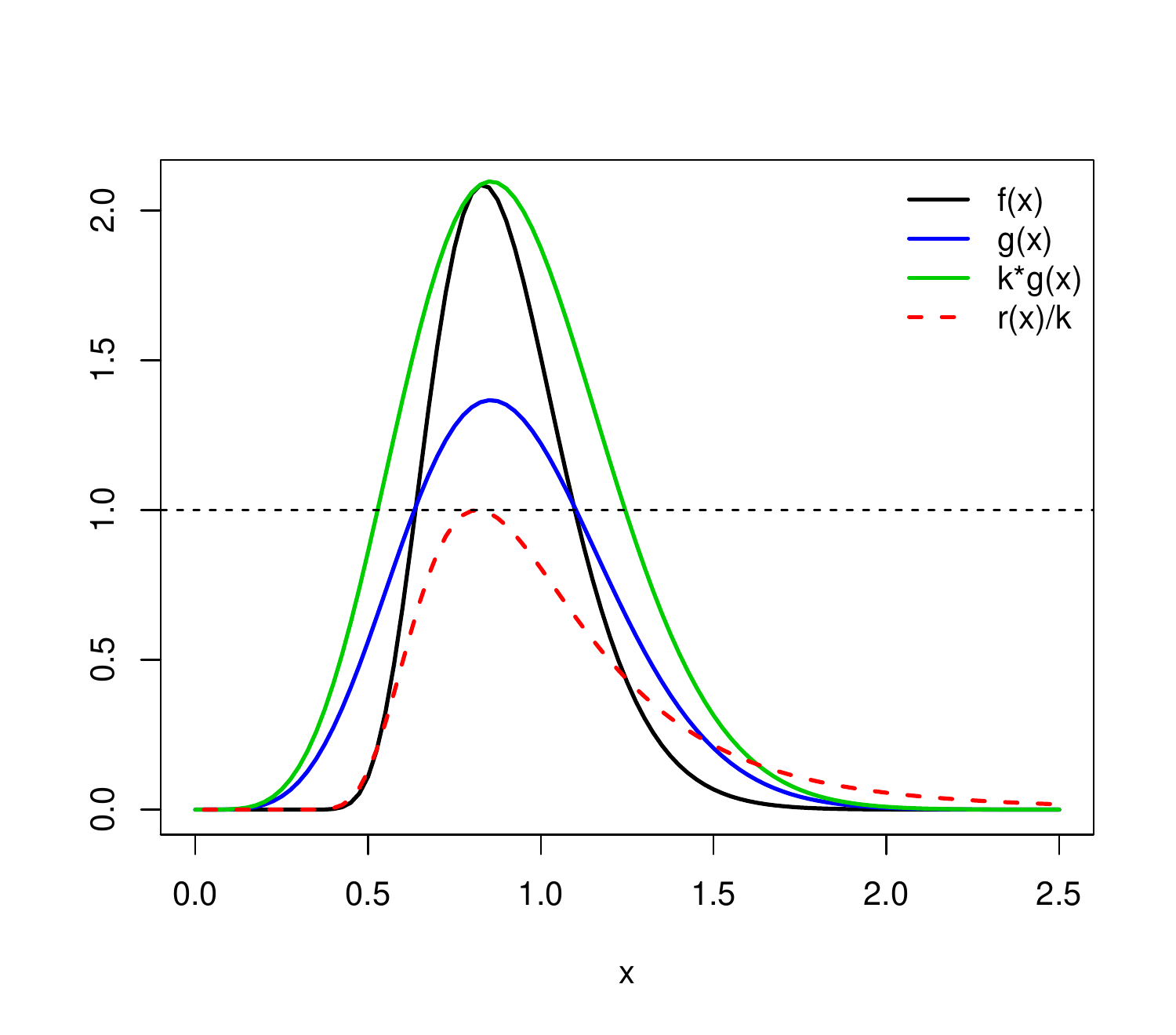}
	\vspace{-0.5cm}
	\caption{Comparison of $f(x)$ and $g(x)$ with $\psi = 2$, $\alpha^2 = 2.25$ and $\beta = -2$. Estimated parameter $\gamma = 5$; the acceptance probability is $a = 0.65$. }
	\label{fig:rejection}
\end{figure}

The best theoretical way to choose the parameter $\gamma$ is to differentiate $k = r(\dot{x}) c_f / c_g$ with respect to $\gamma$ and find the optimal value. However, this can only be done with iterative methods, which negatively affects the computational efficiency of the method. An alternative solution is to set parameter $\gamma$ such that distributions $f$ and $g$ have the maximum at the same value $x_{max}$. That is, the maximum of $f$ is computed by solving a quartic equation and the parameter $\gamma$ is estimated as 
\begin{equation*}
	\gamma = x_{max} \left(2\alpha^2 x_{max} - \beta\right) > -1.
\end{equation*}
Since $\gamma \in \mathbb{N}_0^+$, the closest non-negative integer is chosen. An example of the method is shown in Figure \ref{fig:rejection} of Supporting Information, where the normalizing constant $c_f$ is computed by numerical integration. 

For some settings of the parameters $\psi$, $\alpha$ and $\beta$ the acceptance probability of the proposed algorithm decreases toward $0$, affecting the efficiency of the algorithm. For this reason, and for avoiding the explicit inversion of covariance matrix $\boldsymbol{\Sigma}_\beta^\star$, in the next section we present a Variational Bayes algorithm which overcomes these problems.

\section{Web Appendix C}
\label{sec:C}
\subsection{Variational Inference}
\label{app:variational}

\textit{Variational inference} (VI) is a deterministic optimization method to approximate the target density $\pi\left(\boldsymbol{\theta} \mid \mathbf{y}, \mathbf{X}, \mathbf{Z}\right)$, with $\boldsymbol{\theta} = \left(\boldsymbol{\beta}, \lambda_0^2, \boldsymbol{\lambda}, \psi_0, \boldsymbol{\varphi}, \boldsymbol{\gamma}, \boldsymbol{\kappa}, \tau^2, \zeta, \sigma^2\right)$, with another variational distribution $q\left(\boldsymbol{\theta}\right)$ and reduces Bayesian inference to an optimization problem \citep{Salimans-2015,Lee-2022}. The goal is to find $q(\boldsymbol{\theta})$ that minimizes the Kullback-Leibler divergence (KL) between the target density and the variational distribution. The minimization problem in is eventually reduced to the maximization of the variational lower bound, defined as $\mathcal{L} =  \mathbb{E}_q\left[\log \pi\left(\boldsymbol{\theta}, \mathbf{y}, \mathbf{X}, \mathbf{Z}\right)\right] -  \mathbb{E}_q\left[\log q\left(\boldsymbol{\theta}\right)\right]$. 

A common factorization for $q(\boldsymbol{\theta})$ is the so-called \textit{mean-field Variational approximation} \citep{Jordan-1999, Beal-2003}, which is a compromise between computational tractability and accuracy of the performances. The variational family $q(\boldsymbol{\theta})$ is assumed to be the product of independent marginal variational factors $q_k\left(\theta_k\right)$, $k = 1, \dots, K$. We rely on the \textit{Coordinate Ascent Variational Inference} algorithm (CAVI) \citep{Bishop-2006,Blei-2017} to solve the optimization problem above. Until convergence of the lower bound $\mathcal{L}$, the CAVI algorithm iteratively updates the parameters of the variational factors $q_k\left(\theta_k\right)$, $k = 1, \dots, K$, based on prior distributions' hyperparameters and the current expectation of factor $q_{-k}\left(\theta_{-k}\right)$, considered fixed \citep{Lee-2022}. This way the model is able to account for non-linear dependencies among the parameters. Under the mean field approximation, where the components are assumed to be independent, the optimal solution is given by
\begin{equation*}
	q^\star\left(\theta_k\right) \propto \exp\left\{\mathbb{E}_{q_{-k}}\left[ \log \pi\left(\theta_k \mid \theta_{-k}, \mathbf{y}, \mathbf{X}, \mathbf{Z}\right) \right]\right\}.
\end{equation*}
While the assumption of independence between factors is particularly strict, the CAVI algorithm provides a flexible approach and is ensured to converge to a local optimum \citep{Blei-2017}. Note that, when working with exponential families in a conjugated framework, variational factor $q(\theta_k)$ has the same kernel of the tractable distribution $\pi\left(\theta_k \mid \theta_{-k}, \mathbf{y}, \mathbf{X}, \mathbf{Z}\right)$.

The variational lower bound is computed as
\begin{align}
	\mathcal{L} = &\; \mathbb{E}_{q^\star}\left[\log \pi\left(\boldsymbol{\theta}, \mathbf{y}, \mathbf{X}, \mathbf{Z}\right) \right] - \mathbb{E}_{q^\star}\left[\log q^\star\left(\boldsymbol{\theta}\right) \right] \nonumber \\
	= &\; \mathbb{E}_{q^\star}\left[\log p\left( \mathbf{y} \mid \mathbf{X}, \boldsymbol{\beta}, \sigma^2 \right) \right] + \mathbb{E}_{q^\star}\left[\log \pi\left(\boldsymbol{\beta}\mid \sigma^2, \tau^2, \lambda_0^2, \boldsymbol{\lambda} \right) \right] + \mathbb{E}_{q^\star}\left[\log \pi\left(\lambda_0^2 \mid \psi_0 \right) \right] +   \nonumber \\
	&\; \mathbb{E}_{q^\star}\left[\log \pi\left(\psi_0 \right) \right] + \mathbb{E}_{q^\star}\left[\log \pi\left(\boldsymbol{\lambda} \mid \mathbf{Z}, \boldsymbol{\gamma},\boldsymbol{\varphi}^2 \right) \right] + \mathbb{E}_{q^\star}\left[\log \pi\left(\boldsymbol{\varphi}^2 \right) \right] + \mathbb{E}_{q^\star}\left[\log \pi\left(\boldsymbol{\gamma} \mid \boldsymbol{\kappa}^2\right) \right] +  \nonumber \\
	&\; \mathbb{E}_{q^\star}\left[\log \pi\left(\boldsymbol{\kappa}^2 \right) \right] + \mathbb{E}_{q^\star}\left[\log \pi\left(\tau^2 \mid \zeta \right) \right] + \mathbb{E}_{q^\star}\left[\log \pi\left(\zeta \right) \right] + \mathbb{E}_{q^\star}\left[\log \pi\left(\sigma^2 \right) \right] -  \nonumber \\
	&\; \mathbb{E}_{q^\star}\left[\log q^\star\left(\boldsymbol{\beta} \right) \right] - \mathbb{E}_{q^\star}\left[\log q^\star\left(\lambda_0^2 \right) \right] - \mathbb{E}_{q^\star}\left[\log q^\star\left(\psi_0\right) \right] - \mathbb{E}_{q^\star}\left[\log q^\star\left(\boldsymbol{\lambda} \right) \right] - \mathbb{E}_{q^\star}\left[\log q^\star\left(\boldsymbol{\varphi}^2 \right) \right]  - \nonumber \\
	&\; \mathbb{E}_{q^\star}\left[\log q^\star\left(\boldsymbol{\gamma} \right) \right] - \mathbb{E}_{q^\star}\left[\log q^\star\left(\boldsymbol{\kappa}^2 \right) \right] -  \mathbb{E}_{q^\star}\left[\log q^\star\left(\tau^2 \right) \right] - \mathbb{E}_{q^\star}\left[\log q^\star\left(\zeta \right) \right] - \mathbb{E}_{q^\star}\left[\log q^\star\left(\sigma^2 \right) \right] \nonumber \\
	\propto &\; \frac{1}{2}\left(\log \vert\boldsymbol{\Sigma}_\beta^\star\vert + \log \vert\boldsymbol{\Sigma}_\gamma^\star\vert\right) + \frac{p+1}{2} \mathbb{E}_{\sigma^2}\left[\log\sigma^2\right] - \left(v + \frac{n+p+1}{2} \right) \log l^\star +  \nonumber \\
	& \; \sum_{j=1}^p \left(\log s_j - \log k_j + a_j^\star \mathbb{E}_{\lambda}\left[\lambda_j^{-2}\right] + b_j^\star \mathbb{E}_{\lambda}\left[\lambda_j^2\right] - c_j^\star \mathbb{E}_{\lambda}\left[\lambda_j\right] - \log d_j^\star \right) -  \nonumber \\
	& \;  \sum_{d=1}^D e_d^\star \log f_d^\star - \left(\frac{p}{2} + 1\right) \log g^\star - \log h^\star - \log a_0^\star - \log k_0^\star,
	\label{eq:lb2}
\end{align}
where $k_j = 1 - \mathbb{P}\left(\mathcal{N}\left(\mathbf{z}_j^\intercal \boldsymbol{\gamma}, s_0^2 \varphi_j^2\right) < 0 \right)$ and $s_j$ is the unknown normalizing constant of parameters $\lambda_j$ in \eqref{eq:postlambda_j} of the main article. Each component is derived in Section \ref{app:a1} of Supporting Information.

As opposed to the Gibbs sampler, the Variational algorithm does not require the explicit inversion of the $p \times p$ matrix $\boldsymbol{\Sigma}_\beta^\star$, as only the quantities $\boldsymbol{\Sigma}_\beta^\star \mathbf{X}^\intercal\mathbf{y}$, $\text{diag}\left(\boldsymbol{\Sigma}_\beta^\star\right)$ and $\mathbf{X} \boldsymbol{\Sigma}_\beta^\star \mathbf{X}^\intercal$ are needed. These can be efficiently evaluated with complexity $\mathcal{O}\left(n^2 p\right)$ following the work of \cite{Munch-2019}. The details are shown in Section \ref{app:b1} of Supporting Information. The bottleneck of the algorithm is the computation of the terms $\log s_j$, $\mathbb{E}_\lambda \left[\lambda_j\right]$, $\mathbb{E}_\lambda \left[\lambda_j^2\right]$ and $\mathbb{E}_\lambda \left[\lambda_j^{-2}\right]$, since no closed-form is available. These can be evaluated by numerical integration, for example with the Gaussian quadrature rule or its adaptive variation called Gauss-Kronrod quadrature formula. Depending on the posterior parameters $a_j^\star$, $b_j^\star$ and $c_j^\star$, this strategy is prone to numerical instability. To avoid overflow problem, we evaluate these integrals with the following steps:
\begin{enumerate}
    \item[] \textit{Step 0:} assume we have to evaluate $\int_0^\infty f(x) dx$, where $f(x) \propto x^{\nu} \exp\left\{-d x^{-2} - b x^2 + c x\right\} \cdot \mathbb{I}_{\left(\lambda_j > 0\right)}$, where $\nu = \left\{-3, -1, 0, 1\right\}$;
    \item[] \textit{Step 1:} compute the maximum $\dot{x}$ by solving a quartic equation and evaluate $f(\dot{x})$ on the log-scale;
    \item[] \textit{Step 2:} evaluate $i_0 = \int_0^\infty \exp\left\{\log f(x) - \log f(\dot{x}) \right\}dx$ numerically;
    \item[] \textit{Step 3:} return $\exp\left\{\log i_0 + \log f(\dot{x})\right\}$.
\end{enumerate}

\subsection{Variational Inference in probit regression}
\label{app:probit}

Let $\mathbf{y} \in \left\{0, 1\right\}^n$ be a binary vector and $\mathbf{X} \in \mathbb{R}^{n \times (p+1)}$ be the design matrix. Following \cite{Albert-1993} we introduce the latent variable $\mathbf{w} \in \mathbb{R}^n$ to reach a conjugated framework for updating regression coefficient vector $\boldsymbol{\beta}$. The assumptions of the probit model are
\begin{align*}
    y_i \mid \mathbf{w} = &\; 
    \begin{cases}
      1 \quad \text{if} \; w_i > 0 \\
      0 \quad \text{if} \; w_i \le 0,
    \end{cases}\, \\
    w_i \mid \mathbf{X}, \boldsymbol{\beta} \sim &\; \mathcal{N}\left(\mathbf{x}_i^\intercal \boldsymbol{\beta}, 1\right), \quad i = 1, \dots, n, \\
    \beta_0 \mid \tau^2, \lambda_0^2 \sim & \; \mathcal{N}\left(0, \tau^2 \lambda_0^2 \right) \\
    \beta_j \mid \tau, \lambda_j \sim & \; \mathcal{N}\left(0, \tau^2 \lambda_j^2 \right), \quad j = 1, \dots, p.
\end{align*}
The prior distributions for parameters $\boldsymbol{\lambda}, \boldsymbol{\gamma}$ and $\tau$ are the same of Section \ref{sec:model} and the algorithm follows the same steps of Section \ref{sec:posterior} of the main article. The introduction of the latent variable $\mathbf{w}$ allows to formulate the problem as a normal regression model on the latent variables $w_i$. The normal prior distribution for the regression parameters vector $\boldsymbol{\beta}$ is conjugated to the the normal distribution of $\mathbf{w}$. The joint posterior distribution of the model is
\begin{equation*}
    \pi\left(\boldsymbol{\theta}, \mathbf{w}, \mathbf{y}, \mathbf{X}, \mathbf{Z}\right) \propto p\left(\mathbf{y} \mid \mathbf{w}\right) \cdot \pi\left(\mathbf{w} \mid \mathbf{X}, \boldsymbol{\theta}\right) \cdot \pi \left(\boldsymbol{\theta} \mid \mathbf{Z}\right),
\end{equation*}
where $\boldsymbol{\theta} = \left(\boldsymbol{\beta}, \lambda_0^2, \boldsymbol{\lambda}, \psi_0, \boldsymbol{\varphi}, \boldsymbol{\gamma}, \boldsymbol{\kappa}, \tau^2, \zeta \right)$. Under the mean field approximation, the variational factors for parameters $\mathbf{w}$ and $\boldsymbol{\beta}$ are updated as
\begin{align*}
    q^\star\left(w_i\right) = &\; \begin{cases}
      \mathcal{N}_+\left(\mu_i^\star, 1\right) \quad \text{if} \; y_i = 0 \\
      \mathcal{N}_-\left(\mu_i^\star, 1\right) \quad \text{if} \; y_i = 0,
    \end{cases}\,  \\
    q^\star\left(\boldsymbol{\beta}\right) = &\; \mathcal{N}_p\left(\boldsymbol{\mu}_\beta^\star,  \boldsymbol{\Sigma}_\beta^\star\right), \\ 
    & \; \boldsymbol{\mu}_\beta^\star = \boldsymbol{\Sigma}_\beta^\star \mathbf{X}^\intercal \mathbb{E}_w\left[\mathbf{w}\right], \quad \boldsymbol{\Sigma}_\beta^\star = \left(\mathbf{X}^\intercal \mathbf{X} + \mathbb{E}_{\lambda_0^2 \cdot \lambda \cdot \tau^2} \left[\tau^{-2} \boldsymbol{\Lambda}^{-2}\right]\right)^{-1},  \nonumber
\end{align*}
where $\mu_i^\star = \mathbf{x}_i^\intercal \mathbb{E}_\beta\left[\boldsymbol{\beta}\right]$ and $\mathcal{N}_+$ and $\mathcal{N}_-$ denote a normal distribution left and right truncated at 0, respectively. The expectation of latent variables $w_i$ is
\begin{equation*}
    \mathbb{E}_w\left[w_i\right] = \begin{cases}
      \mu_i^\star + \frac{\phi_i}{1 - \Phi_i} \quad \text{if} \; y_i = 0 \\
      \mu_i^\star - \frac{\phi_i}{\Phi_i} \quad \quad \text{if} \; y_i = 0,
    \end{cases}\,
\end{equation*}
where $\phi_i = \phi\left(-\mu_i^\star \right)$ and $\Phi_i = \Phi\left(-\mu_i^\star \right)$ are the normal density and the normal cumulative density functions, respectively. The variational lower bound is computed as
\begin{align}
	\mathcal{L} = &\; \mathbb{E}_{q^\star}\left[\log \pi\left(\boldsymbol{\theta}, \mathbf{w}, \mathbf{y}, \mathbf{X}, \mathbf{Z}\right) \right] - \mathbb{E}_{q^\star}\left[\log q^\star\left(\boldsymbol{\theta}, \mathbf{w}\right) \right] \nonumber \\
	= &\; \mathbb{E}_{q^\star}\left[\log p\left( \mathbf{y}\mid \mathbf{w}\right) \right] + \mathbb{E}_{q^\star}\left[\log \pi \left( \mathbf{w} \mid \mathbf{X}, \boldsymbol{\beta} \right) \right] + \mathbb{E}_{q^\star}\left[\log \pi\left(\boldsymbol{\beta}\mid \tau^2, \lambda_0^2, \boldsymbol{\lambda} \right) \right] + \mathbb{E}_{q^\star}\left[\log \pi\left(\lambda_0^2 \mid \psi_0 \right) \right] +   \nonumber \\
	&\; \mathbb{E}_{q^\star}\left[\log \pi\left(\psi_0 \right) \right] +  \mathbb{E}_{q^\star}\left[\log \pi\left(\boldsymbol{\lambda} \mid   \mathbf{Z}, \boldsymbol{\gamma},\boldsymbol{\varphi}^2 \right) \right] + \mathbb{E}_{q^\star}\left[\log \pi\left(\boldsymbol{\varphi}^2 \right) \right] + \mathbb{E}_{q^\star}\left[\log \pi\left(\boldsymbol{\gamma} \mid \boldsymbol{\kappa}^2\right) \right] +   \nonumber \\
	&\; \mathbb{E}_{q^\star}\left[\log \pi\left(\boldsymbol{\kappa}^2 \right) \right] + \mathbb{E}_{q^\star}\left[\log \pi\left(\tau^2 \mid \zeta \right) \right] + \mathbb{E}_{q^\star}\left[\log \pi\left(\zeta \right) \right] - \mathbb{E}_{q^\star}\left[\log q^\star\left(\mathbf{w} \right) \right] - \mathbb{E}_{q^\star}\left[\log q^\star\left(\boldsymbol{\beta} \right) \right] - \nonumber \\
	&\; \mathbb{E}_{q^\star}\left[\log q^\star\left(\lambda_0^2 \right) \right] - \mathbb{E}_{q^\star}\left[\log q^\star\left(\psi_0\right)\right] - \mathbb{E}_{q^\star}\left[\log q^\star\left(\boldsymbol{\lambda} \right) \right] - \mathbb{E}_{q^\star}\left[\log q^\star\left(\boldsymbol{\varphi}^2 \right) \right]  - \mathbb{E}_{q^\star}\left[\log q^\star\left(\boldsymbol{\gamma} \right) \right] - \nonumber \\
	&\; \mathbb{E}_{q^\star}\left[\log q^\star\left(\boldsymbol{\kappa}^2 \right) \right] - \mathbb{E}_{q^\star}\left[\log q^\star\left(\tau^2 \right) \right] - \mathbb{E}_{q^\star}\left[\log q^\star\left(\zeta \right) \right]  \nonumber \\
	\propto &\; \sum_{i = 1}^n \left(y_i \log\left(1 - \Phi_i\right) + (1- y_i) \log\left(\Phi_i\right)\right) + \frac{1}{2}\left(\log \vert\boldsymbol{\Sigma}_\beta^\star\vert + \log \vert\boldsymbol{\Sigma}_\gamma^\star\vert\right) +  \nonumber \\
	& \;  \sum_{j=1}^p \left(\log s_j - \log k_j\right) + \sum_{j=1}^p \left(a_j^\star \mathbb{E}_{\lambda}\left[\lambda_j^{-2}\right] + b_j^\star \mathbb{E}_{\lambda}\left[\lambda_j^2\right] - c_j^\star \mathbb{E}_{\lambda}\left[\lambda_j\right] - \log d_j^\star \right) -  \nonumber \\
	& \;  \sum_{d=1}^D e_d^\star \log f_d^\star - \left(\frac{p}{2} + 1\right) \log g^\star - \log h^\star - \log a_0^\star - \log k_0^\star,
	\label{eq:lb_probit}
\end{align}
where all the quantities are defined in Section \ref{app:a2} of Supporting Information. The method requires the computation of $\boldsymbol{\Sigma}_\beta^\star \mathbf{X}^\intercal \mathbb{E}_w\left[w\right]$ and $\text{diag}\left(\boldsymbol{\Sigma}_\beta^\star\right)$, which can efficiently be evaluated by applying the strategy in Section \ref{app:b1} of Supporting Information.

\section{Web Appendix D}
\label{sec:D}

Here we report the components for the evaluation of the lower bound $\mathcal{L}$ in \eqref{eq:lb2} and \eqref{eq:lb_probit} of Supporting Information. The results below rely on the property $\mathbb{E}_x \left[\mathbf{x}^\intercal \mathbf{A} \mathbf{x} \right] = \mathbb{E}_x (\mathbf{x})^\intercal \mathbf{A} \mathbb{E}_x (\mathbf{x}) + tr\left(\mathbf{A} \mathbf{V} \right) $, where $\mathbf{V}$ is the correlation matrix of generical random variable $\mathbf{x}$. 

\subsection{Linear regression}
\label{app:a1}

The variational lower bound $\mathcal{L}$ in \eqref{eq:lb2} is computed with the following components:

\begin{align*}
    & \mathbb{E}_{q^\star}\left[\log p\left(\mathbf{y} \mid \boldsymbol{\beta}, \sigma^2, \boldsymbol{\lambda} \right) \right] \propto -\frac{n}{2} \mathbb{E}_{\sigma^2}\left[\log \sigma^2\right] - \frac{1}{2}\mathbb{E}_{\beta \cdot \sigma^2}\left[\frac{\Vert \mathbf{y} - \mathbf{X}\boldsymbol{\beta}\Vert_2^2}{\sigma^2}\right] \\
    & \mathbb{E}_{q^\star}\left[\log \pi\left(\boldsymbol{\beta}\mid \sigma^2, \tau^2, \lambda_0^2, \boldsymbol{\lambda} \right) \right] \propto - \frac{p+1}{2}\mathbb{E}_{\sigma^2} \left[ \log \sigma^2\right] - \frac{p+1}{2}\mathbb{E}_{\tau^2} \left[ \log \tau^2\right] - \frac{1}{2}\mathbb{E}_{\lambda_0^2} \left[\log \lambda_0^2\right] - \\ & \hspace{5.3cm} \sum_{j=1}^p\mathbb{E}_\lambda \left[\log \lambda_j\right] - \frac{1}{2} \sum_{j=0}^p\mathbb{E}_{\beta_0 \cdot \beta  \cdot \lambda_0^2 \cdot \lambda \cdot \tau^2 \cdot \sigma^2}\left[\frac{\beta_j^2}{\sigma^2 \tau^2 \lambda_j^2 }\right]   \\
    & \mathbb{E}_{q^\star}\left[\log \pi\left(\lambda_0^2 \mid \psi_0 \right) \right] \propto  -\frac{1}{2}\mathbb{E}_{\psi_0} \left[\log\psi_0 \right]  - \frac{3}{2} \mathbb{E}_{\lambda_0^2} \left[\log \lambda_0^2\right] - \mathbb{E}_{\lambda_0^2 \cdot \psi_0} \left[\frac{1}{\psi_0 \lambda_0^2}\right]  \\
    & \mathbb{E}_{q^\star}\left[\log \pi\left(\psi_0 \right) \right] \propto -\frac{3}{2}\mathbb{E}_{\psi_0} \left[\log \psi_0\right] - \mathbb{E}_{\psi_0} \left[ \psi_0^{-1}\right]   \\
    & \mathbb{E}_{q^\star}\left[\log \pi\left(\boldsymbol{\lambda} \mid \boldsymbol{\gamma},\boldsymbol{\varphi}^2 \right) \right] \propto  \sum_{j = 1}^p \Bigg(- \log k_j - \frac{1}{2} \mathbb{E}_{\varphi^2}\left[\log \varphi_j^2\right] - \frac{1}{2s_0^2} \mathbb{E}_{\lambda \cdot \gamma \cdot \varphi^2} \left[\frac{\left(\lambda_j - \mathbf{z}_j^\intercal \boldsymbol{\gamma}\right)^2}{\varphi_j^2}\right] \Bigg) \\
    & \mathbb{E}_{q^\star}\left[\log \pi\left(\boldsymbol{\varphi}^2 \right) \right] \propto -\frac {1}{2}\sum_{j=1}^p \Bigg(3\mathbb{E}_{\varphi^2} \left[ \log \varphi_j^2\right] + \mathbb{E}_{\varphi^2} \left[ \varphi_j^{-2}\right] \Bigg) \\
    & \mathbb{E}_{q^\star}\left[\log \pi\left(\boldsymbol{\gamma} \mid \boldsymbol{\kappa}^2\right) \right] \propto - \frac{1}{2} \sum_{d = 1}^D \Bigg( m_d \mathbb{E}_{\kappa^2} \left[\log\kappa_d^2\right]  + \mathbb{E}_{\gamma \cdot \kappa^2} \left[ \frac{\boldsymbol{\gamma}_d^\intercal \boldsymbol{\gamma}_d}{\kappa_d^2}\right]\Bigg)  \\
    & \mathbb{E}_{q^\star}\left[\log \pi\left(\boldsymbol{\kappa}^2 \right) \right] \propto - \sum_{d = 1}^D \Bigg((a_d +1) \mathbb{E}_{\kappa^2} \left[ \log \kappa_d^2\right] + b_d \mathbb{E}_{\kappa^2} \left[ \kappa_d^{-2}\right] \Bigg) \\
    & \mathbb{E}_{q^\star}\left[\log \pi\left(\tau^2 \mid \zeta \right) \right] \propto -\frac{1}{2}\mathbb{E}_\zeta \left[\log\zeta \right]  - \frac{3}{2} \mathbb{E}_{\tau^2} \left[\log \tau^2\right] - \mathbb{E}_{\tau^2 \cdot \zeta} \left[\frac{1}{\zeta \tau^2}\right]  \\ 
    & \mathbb{E}_{q^\star}\left[\log \pi\left(\zeta \right) \right] \propto -\frac{3}{2}\mathbb{E}_\zeta \left[\log \zeta\right] - \mathbb{E}_\zeta \left[ \zeta^{-1}\right]  \\
    & \mathbb{E}_{q^\star}\left[\log \pi\left(\sigma^2\right) \right] \propto -\left(v + 1\right) \mathbb{E}_{\sigma^2}\left[\log \sigma^2\right] - q\mathbb{E}_{\sigma^2}\left[\sigma^{-2}\right]  \\
    & \mathbb{E}_{q^\star}\left[\log q^\star\left(\boldsymbol{\beta} \right) \right] \propto -\frac{p+1}{2} \mathbb{E}_{\sigma^2}\left[\log \sigma^2\right] - \frac{1}{2} \log\vert\boldsymbol{\Sigma}_\beta^\star\vert - \frac{p+1}{2} \\
    & \mathbb{E}_{q^\star}\left[\log q^\star\left(\lambda_0^2 \right) \right] \propto \log\left(a_0^\star\right) - 2 \mathbb{E}_{\lambda_0^2}\left[\log \lambda_0^2\right] -  \mathbb{E}_{\lambda_0^2 \cdot \psi_0}\left[\frac{1}{\psi_0 \lambda_0^2}\right] - \frac{1}{2}\mathbb{E}_{\beta_0 \cdot \lambda_0^2 \cdot \tau^2 \cdot \sigma^2}\left[\frac{\beta_0^2}{\sigma^2 \tau^2 \lambda_0^2}\right] \\
    & \mathbb{E}_{q^\star}\left[\log q^\star\left(\psi_0 \right) \right] \propto  \log k_0^\star - 2 \mathbb{E}_{\psi_0} \left[ \log \psi_0\right] - \mathbb{E}_{\psi_0} \left[ \psi_0^{-1}\right] - \mathbb{E}_{\lambda_0^2 \cdot \psi_0} \left[\frac{1}{\psi_0 \lambda_0^2}\right]
    \end{align*}
    \begin{align*}
    & \mathbb{E}_{q^\star}\left[\log q^\star\left(\boldsymbol{\lambda} \right) \right] \propto  \sum_{j=1}^p\Bigg(-\log s_j -  \mathbb{E}_\lambda \left[\log \lambda_j\right] - \frac{1}{2}\mathbb{E}_{\beta \cdot \lambda \cdot \sigma^2 \cdot \tau^2}\left[\frac{\beta_j^2}{\sigma^2 \tau^2 \lambda_j^2}\right] - \\ & \hspace{3cm} \frac{1}{2s_0^2} \mathbb{E}_{\lambda \cdot \varphi^2} \left[\frac{\lambda_j^2}{\varphi_j^2}\right] +  \frac{1}{s_0^2}\mathbf{z}_j^\intercal \mathbb{E}_{\lambda \cdot \gamma \cdot \varphi^2} \left[\frac {\boldsymbol{\gamma}\lambda_j}{\varphi_j^2}\right] \Bigg) \\
    & \mathbb{E}_{q^\star}\left[\log q^\star\left(\boldsymbol{\varphi}^2 \right) \right] \propto \sum_{j=1}^p \Bigg(\log d_j^\star - 2 \mathbb{E}_{\varphi^2} \left[ \log \varphi_j^2\right] -  \frac{1}{2}\mathbb{E}_{\varphi^2} \left[ \varphi_j^{-2}\right] - \frac{1}{2s_0^2} \mathbb{E}_{\lambda \cdot \gamma \cdot \varphi^2} \left[ \frac{\left(\lambda_j - \mathbf{z}_j^\intercal \boldsymbol{\gamma}\right)^2}{\varphi_j^2}\right]\Bigg) \\
    & \mathbb{E}_{q^\star}\left[\log q^\star\left(\boldsymbol{\gamma} \right) \right] \propto - \frac{1}{2} \log\vert\boldsymbol{\Sigma}_\gamma^\star\vert - \frac{M}{2s_0^2} \\     
    & \mathbb{E}_{q^\star}\left[\log q^\star\left(\kappa^2 \right) \right] \propto \sum_{d=1}^D \Bigg(\left(a_d + \frac{m_d}{2}\right) \log f_d^\star - b_d \mathbb{E}_{\kappa^2}\left[\kappa_d^{-2}\right] - \left(a_d + \frac{m_d}{2} +1 \right) \mathbb{E}_{\kappa^2}\left[\log \kappa_d^2\right] - \\ & \hspace{3.7cm} \frac{1}{2}\mathbb{E}_{\gamma \cdot \kappa^2}\left[\frac{\boldsymbol{\gamma}_d^\intercal \boldsymbol{\gamma}_d}{\kappa_d^2}\right] \Bigg) \\
    & \mathbb{E}_{q^\star}\left[\log q^\star\left(\tau^2 \right) \right] \propto  \left(\frac{p}{2}+1\right) \log g^\star - \left(\frac{p}{2} + 2\right) \mathbb{E}_{\tau^2}\left[\log \tau^2\right] -  \mathbb{E}_{\tau^2 \cdot \zeta }\left[\frac{1}{\zeta\tau^2}\right] - \\ & \hspace{3.2cm} \frac{1}{2} \sum_{j=0}^p \mathbb{E}_{\beta_0 \cdot \beta \cdot \lambda_0^2 \cdot \lambda \cdot \tau^2 \cdot \sigma^2 }\left[\frac{\beta_j^2}{\sigma^2 \tau^2 \lambda_j^2}\right]  \\ 
    & \mathbb{E}_{q^\star}\left[\log q^\star\left(\zeta \right) \right] \propto \log h^\star - 2 \mathbb{E}_\zeta \left[ \log \zeta\right] - \mathbb{E}_\zeta \left[ \zeta^{-1}\right] - \mathbb{E}_{\tau^2 \cdot \zeta} \left[\frac{1}{\zeta \tau^2}\right]  \\
    & \mathbb{E}_{q^\star}\left[\log q^\star\left(\sigma^2 \right) \right] \propto \left(v + \frac{n+p+1}{2}\right) \log l^\star -  q\mathbb{E}_{\sigma^2}\left[\sigma^{-2}\right] - \left(v + \frac{n+p+3}{2}\right)\mathbb{E}_{\sigma^2}\left[\log \sigma^2\right] - \\ & \hspace{3.5cm} \frac{1}{2}\mathbb{E}_{\beta \cdot \sigma^2}\left[\frac{\Vert\mathbf{y} - \mathbf{X}\boldsymbol{\beta}\Vert_2^2}{\sigma^2}\right] - \frac{1}{2}\sum_{j=0}^p\mathbb{E}_{\beta_0 \cdot \beta \cdot \lambda_0^2 \cdot \lambda \cdot \sigma^2 \cdot \tau^2}\left[\frac{\beta_j^2}{\sigma^2 \tau^2 \lambda_j^2}\right] 
\end{align*}

\subsection{Probit regression}
\label{app:a2}

The variational lower bound $\mathcal{L}$ in \eqref{eq:lb_probit} of Supporting Information is computed with the following components:
\begin{align*}
    & \mathbb{E}_{q^\star}\left[\log p\left(\mathbf{y} \mid \mathbf{w} \right) \right] \propto \sum_{i = 1}^n \Bigg( \mathbb{E}_w\left[y_i \log(\mathbb{I}_{\left(w_i > 0\right)}\right] + \mathbb{E}_w\left[(1 - y_i) \log(\mathbb{I}_{\left(w_i < 0\right)}\right] \Bigg) = 0 \\
    & \mathbb{E}_{q^\star}\left[\log p\left(\mathbf{w} \mid \boldsymbol{\beta} \right) \right] \propto -\frac{1}{2} \sum_{i = 1}^n \mathbb{E}_w\left[\Vert w_i - \mathbf{x}_i^\intercal \boldsymbol{\beta}\Vert_2^2\right]  \\
    & \mathbb{E}_{q^\star}\left[\log \pi\left(\boldsymbol{\beta}\mid \sigma^2, \tau^2, \lambda_0^2, \boldsymbol{\lambda} \right) \right] \propto -\frac{p+1}{2}\mathbb{E}_{\tau^2} \left[ \log \tau^2\right] - \frac{1}{2}\mathbb{E}_{\lambda_0^2} \left[\log \lambda_0^2\right] -  \sum_{j=1}^p\mathbb{E}_\lambda \left[\log \lambda_j\right] -  \\ &\hspace{5.6cm} \frac{1}{2} \sum_{j=0}^p\mathbb{E}_{\beta_0 \cdot \beta \cdot \lambda_0^2 \cdot \lambda \cdot \tau^2}\left[\frac{\beta_j^2}{\tau^2 \lambda_j^2 }\right]   \\
    & \mathbb{E}_{q^\star}\left[\log \pi\left(\lambda_0^2 \mid \psi_0 \right) \right] \propto -\frac{1}{2}\mathbb{E}_{\psi_0} \left[\log\psi_0 \right]  - \frac{3}{2} \mathbb{E}_{\lambda_0^2} \left[\log \lambda_0^2\right] - \mathbb{E}_{\lambda_0^2 \cdot \psi_0} \left[\frac{1}{\psi_0 \lambda_0^2}\right]  \\
    & \mathbb{E}_{q^\star}\left[\log \pi\left(\psi_0 \right) \right] \propto -\frac{3}{2}\mathbb{E}_{\psi_0} \left[\log \psi_0\right] - \mathbb{E}_{\psi_0} \left[ \psi_0^{-1}\right]   \\
    & \mathbb{E}_{q^\star}\left[\log \pi\left(\boldsymbol{\lambda} \mid \boldsymbol{\gamma},\boldsymbol{\varphi}^2 \right) \right] \propto \sum_{j = 1}^p \Bigg(-\log k_j - \frac{1}{2} \mathbb{E}_{\varphi^2}\left[\log \varphi_j^2\right] -\frac{1}{2s_0^2} \mathbb{E}_{\lambda \cdot \gamma \cdot \varphi^2} \left[\frac{\left(\lambda_j - \mathbf{z}_j^\intercal \boldsymbol{\gamma}\right)^2}{\varphi_j^2}\right] \Bigg)  \\
    & \mathbb{E}_{q^\star}\left[\log \pi\left(\boldsymbol{\varphi}^2 \right) \right] \propto  - \frac{1}{2} \sum_{j=1}^p \Bigg( 3\mathbb{E}_{\varphi^2} \left[ \log \varphi_j^2\right] + \mathbb{E}_{\varphi^2} \left[ \varphi_j^{-2}\right] \Bigg) \\
    & \mathbb{E}_{q^\star}\left[\log \pi\left(\boldsymbol{\gamma} \mid \boldsymbol{\kappa}^2\right) \right] \propto  - \frac{1}{2} \sum_{d = 1}^D \Bigg( m_d \mathbb{E}_\kappa \left[\log\kappa_d^2\right] +  \mathbb{E}_{\gamma \cdot \kappa} \left[ \frac{\boldsymbol{\gamma}_d^\intercal \boldsymbol{\gamma}_d}{\kappa_d^2}\right]  \Bigg) \\
    & \mathbb{E}_{q^\star}\left[\log \pi\left(\boldsymbol{\kappa}^2 \right) \right] \propto  \sum_{d = 1}^D \Bigg(-(a_d +1) \mathbb{E}_{\kappa^2} \left[ \log \kappa_d^2\right]- b_d \mathbb{E}_{\kappa^2} \left[ \kappa_d^{-2}\right] \Bigg) \\
    & \mathbb{E}_{q^\star}\left[\log \pi\left(\tau^2 \mid \zeta \right) \right] \propto -\frac{1}{2}\mathbb{E}_\zeta \left[\log\zeta \right]  - \frac{3}{2} \mathbb{E}_{\tau^2} \left[\log \tau^2\right] - \mathbb{E}_{\tau^2 \cdot \zeta} \left[\frac{1}{\zeta \tau^2}\right]  \\ 
    & \mathbb{E}_{q^\star}\left[\log \pi\left(\zeta \right) \right] \propto -\frac{3}{2}\mathbb{E}_\zeta \left[\log \zeta\right] - \mathbb{E}_\zeta \left[ \zeta^{-1}\right]  \\
    & \mathbb{E}_{q^\star}\left[\log q^\star\left(\mathbf{w} \right) \right] \propto \sum_{i = 1}^n \Bigg( -\frac{1}{2}  \mathbb{E}_w\left[\Vert w_i - \mathbf{x}_i^\intercal \boldsymbol{\beta}\Vert_2^2\right] -  y_i \log\left(1 - \Phi_i\right) - \left(1 - y_i\right) \log\left(\Phi_i\right)\Bigg)  \\
    & \mathbb{E}_{q^\star}\left[\log q^\star\left(\boldsymbol{\beta} \right) \right] \propto -\frac{1}{2}\log\vert\boldsymbol{\Sigma}_\beta^\star\vert - \frac{p+1}{2}  \\
    & \mathbb{E}_{q^\star}\left[\log q^\star\left(\lambda_0^2 \right) \right] \propto \log\left(a_0^\star\right) - 2 \mathbb{E}_{\lambda_0^2}\left[\log \lambda_0^2\right] -  \mathbb{E}_{\lambda_0^2 \cdot \psi_0}\left[\frac{1}{\psi_0 \lambda_0^2}\right] - \frac{1}{2}\mathbb{E}_{\beta_0 \cdot \tau^2 \cdot \lambda_0^2}\left[\frac{\beta_0^2}{\tau^2 \lambda_0^2}\right] 
    \end{align*}
    \begin{align*}
    & \mathbb{E}_{q^\star}\left[\log q^\star\left(\psi_0 \right) \right] \propto  \log k_0^\star - 2 \mathbb{E}_{\psi_0} \left[ \log \psi_0\right] - \mathbb{E}_{\psi_0} \left[ \psi_0^{-1}\right] - \mathbb{E}_{\lambda_0^2 \cdot \psi_0} \left[\frac{1}{\psi_0 \lambda_0^2}\right] \\
    & \mathbb{E}_{q^\star}\left[\log q^\star\left(\boldsymbol{\lambda} \right) \right] \propto  \sum_{j=1}^p \Bigg(-\log s_j -  \mathbb{E}_\lambda \left[\log \lambda_j\right] - \frac{1}{2}\mathbb{E}_{\beta \cdot \tau^2 \cdot \lambda}\left[\frac{\beta_j^2}{\tau^2 \lambda_j^2}\right] -  \frac {1}{2s_0^2} \mathbb{E}_{\lambda \cdot \varphi^2} \left[\frac{\lambda_j^2}{\varphi_j^2}\right] + \\ & \hspace{3.1cm} \frac{1}{s_0^2} \mathbf{z}_j^\intercal \mathbb{E}_{\lambda \cdot \gamma \cdot \varphi^2} \left[\frac {\boldsymbol{\gamma}\lambda_j}{\varphi_j^2}\right] \Bigg) \\
    & \mathbb{E}_{q^\star}\left[\log q^\star\left(\boldsymbol{\varphi}^2 \right) \right] \propto  \sum_{j=1}^p \Bigg( \log d_j^\star - 2 \mathbb{E}_{\varphi^2} \left[ \log \varphi_j^2\right] - \frac{1}{2} \mathbb{E}_{\varphi^2} \left[ \varphi_j^{-2}\right] - \frac{1}{2s_0^2} \mathbb{E}_{\lambda \cdot \gamma \cdot \varphi^2} \left[ \frac{\left(\lambda_j - \mathbf{z}_j^\intercal \boldsymbol{\gamma}\right)^2}{\varphi_j^2}\right]\Bigg)  \\
    & \mathbb{E}_{q^\star}\left[\log q^\star\left(\boldsymbol{\gamma} \right) \right] \propto - \frac{1}{2} \log\vert\boldsymbol{\Sigma}_\gamma^\star\vert - \frac{M}{2s_0^2}  \\
    & \mathbb{E}_{q^\star}\left[\log q^\star\left(\kappa^2 \right) \right] \propto  \sum_{d=1}^D \Bigg(\left(a_d + \frac{m_d}{2}\right) \log f_d^\star - (a_d + \frac{m_d}{2} +1 ) \mathbb{E}_{\kappa^2}\left[\log \kappa_d^2\right] -   b_d \mathbb{E}_{\kappa^2}\left[\kappa_d^{-2}\right] - \\ & \hspace{3.4cm} \frac{1}{2}\mathbb{E}_{\gamma \cdot \kappa^2}\left[\frac{\boldsymbol{\gamma}_d^\intercal \boldsymbol{\gamma}_d}{\kappa_d^2}\right] \Bigg) \\
    & \mathbb{E}_{q^\star}\left[\log q^\star\left(\tau^2 \right) \right] \propto  \left(\frac{p}{2}+1\right) \log\left(g^\star\right) - \left(\frac{p}{2}+2\right) \mathbb{E}_{\tau^2}\left[\log \tau^2\right] -  \mathbb{E}_{\tau^2}\left[\frac{1}{\zeta\tau^2}\right] - \frac{1}{2} \sum_{j=0}^p \mathbb{E}_{\beta_0 \cdot \beta \cdot \lambda_0^2 \cdot \lambda \cdot \tau^2}\left[\frac{\beta_j^2}{\tau^2 \lambda_j^2}\right]  \\
    & \mathbb{E}_{q^\star}\left[\log q^\star\left(\zeta \right) \right] \propto \log h^\star - 2\mathbb{E}_\zeta \left[ \log \zeta\right] - \mathbb{E}_\zeta \left[ \zeta^{-1}\right] - \mathbb{E}_\zeta \left[\frac{1}{\zeta \tau^2}\right], 
\end{align*}
where $\Phi_i = \Phi\left(-\mathbf{x}_i^\intercal \mathbb{E}_\beta\left[ \boldsymbol{\beta}\right] \right)$.

\newpage
\section{Web Appendix E}
\label{sec:E}
\subsection{Computational methods for the variational algorithm}
\label{app:b1}

The Variational algorithm has the advantage of not requiring the explicit inversion of covariance matrix $\boldsymbol{\Sigma}_\beta^\star = \left(\mathbf{X}^\intercal \mathbf{X} + \boldsymbol{\Delta}\right)^{-1}$, where  $\boldsymbol{\Delta} = \mathbb{E}_{\lambda \cdot \tau^2}\left[\tau^{-2} \Lambda^{-2}\right]$ is diagonal, which becomes computationally infeasible when the number of covariates $p$ increases. The quantities needed are the diagonal entries of $\boldsymbol{\Sigma}_\beta^\star$, the posterior mean $\boldsymbol{\mu}_\beta^\star = \boldsymbol{\Sigma}_\beta^\star \left(\mathbf{X}^\intercal \mathbf{y}\right)$ and $tr\left(\mathbf{X}^\intercal \mathbf{X} \boldsymbol{\Sigma}_\beta^\star\right)$. By means of the Woodbury identity, the needed quantities can be efficiently computed as
\begin{align}
    \text{diag}\left(\boldsymbol{\Sigma}_\beta^\star\right) = &\; \text{diag}\left(\boldsymbol{\Delta}^{-1}\right) - \text{diag}\left(\boldsymbol{\Delta}^{-1} \mathbf{X}^\intercal \left(\mathbf{I}_n + \mathbf{X} \boldsymbol{\Delta}^{-1} \mathbf{X}^\intercal\right)^{-1} \mathbf{X} \boldsymbol{\Delta}^{-1} \right)  \nonumber\\
    = &\; \text{diag}\left(\boldsymbol{\Delta}^{-1}\right) - \left[\left(\boldsymbol{\Delta}^{-1} \mathbf{X}^\intercal\right) \left(\mathbf{I}_n + \mathbf{X} \boldsymbol{\Delta}^{-1} \mathbf{X}^\intercal\right)^{-1} \circ \left(\boldsymbol{\Delta}^{-1} \mathbf{X}^\intercal\right) \right] \cdot \mathbf{1}_{n \times 1},  \label{eq:diag} \\
    \boldsymbol{\mu}_\beta^\star = & \; \left(\boldsymbol{\Sigma}_\beta^\star \mathbf{X}^\intercal \right)\mathbf{y} \nonumber \\
    = & \; \left[\left(\boldsymbol{\Delta}^{-1} \mathbf{X}^\intercal\right) - \left( \boldsymbol{\Delta}^{-1} \mathbf{X}^\intercal \right) \left(\mathbf{I}_n + \mathbf{X} \boldsymbol{\Delta}^{-1} \mathbf{X}^\intercal\right)^{-1} \left(\mathbf{X} \boldsymbol{\Delta}^{-1} \mathbf{X}^\intercal \right) \right] \mathbf{y}, \\
    tr\left( \mathbf{X} \boldsymbol{\Sigma}_\beta^\star \mathbf{X}^\intercal \right) = & \; \sum_{i = 1}^n \mathbf{x}_i^\intercal \boldsymbol{\Sigma}_\beta^\star \mathbf{x}_i = \sum_{i = 1}^n \mathbf{x}_i^\intercal \left(\boldsymbol{\Sigma}_\beta^\star \mathbf{x}_i\right), \label{eq:xSx}
\end{align}
where $\circ$ denotes the Hadamard product, $\mathbf{I}_n$ is the $n\times n$ identity matrix and $\mathbf{1}_{n \times 1}$ denotes the $n$-dimensional vector of ones. In \eqref{eq:xSx} the quantity $\boldsymbol{\Sigma}_\beta^\star \mathbf{x}_i$ has already been computed when evaluating vector $\boldsymbol{\mu}_\beta^\star$ and all the matrix products in \eqref{eq:diag}-\eqref{eq:xSx} can be evaluated with $\mathcal{O}\left(n^2p\right)$ operations, which is linear in $p$.
Note that in the probit regression the last quantity \eqref{eq:xSx} is not needed, thus the posterior mean can be efficiently computed as
\begin{equation*}
    \boldsymbol{\mu}_\beta^\star = \boldsymbol{\Delta}^{-1}\left( \mathbf{X}^\intercal \mathbb{E}_w\left[\mathbf{w}\right] \right) - \left( \boldsymbol{\Delta}^{-1} \mathbf{X}^\intercal \right) \left(\mathbf{I}_n + \mathbf{X} \boldsymbol{\Delta}^{-1} \mathbf{X}^\intercal\right)^{-1} \left(\mathbf{X} \boldsymbol{\Delta}^{-1} \right) \left( \mathbf{X}^\intercal \mathbb{E}_w\left[\mathbf{w}\right] \right),
\end{equation*}
which only involves matrix-vector products, further improving the computational efficiency of the algorithm. Finally, the determinant can be efficiently evaluated as
\begin{equation*}
    \left|\boldsymbol{\Sigma}_\beta^\star\right| = \left|\boldsymbol{\Delta}\right|  / \left|\mathbf{I}_n + \mathbf{X} \boldsymbol{\Delta}^{-1} \mathbf{X}^\intercal\right|.
\end{equation*}

\section{Web Appendix F}
\label{sec:F}
\subsection{Details on the simulation studies}
\label{sec:sim_scheme}

For all the considered cases we rely on the following simulation scheme. Let $\boldsymbol{\beta}^0$ be the true $(p+1)$-dimensional regression parameter vector. We set the number of true non-zero coefficients to $p_0$ (intercept excluded). The entries of design matrices are sampled independently as $x_{ij} \sim \mathcal{N}(0, 1)$, $i = 1, \dots, n$, $j = 1, \dots, p$, whereas the response variable $\mathbf{y}$ and the $(p+1)$-dimensional true regression coefficient vector are sampled following a modified version of the scheme in \citet{Johnson-2013}. Specifically
\newpage
\begin{align*}
    y_i = &\; \mathbf{x}_i^\intercal \boldsymbol{\beta}^0 + \varepsilon_i, \quad \varepsilon_i \sim \mathcal{N}(0, 1), \quad i = 1, \dots, n, \\
    \beta_j^0 = &\;\begin{cases}
                    v_0 \vert t\vert \quad \text{if} \mbox{ } j = 0,\\
                    \left(-1\right)^u \left(v^2 \log(n) / \sqrt{n} + v \vert t \vert\right) \quad \text{if} \mbox{ } j = 1, \dots, p_0, \\
                    0 \quad \text{otherwise},
                \end{cases}
\end{align*}
where $v_0^2 = 0.5$, $v^2 = 0.75$, $u \sim \mathcal{B}(0.4)$ and $t \sim \mathcal{N}\left(0, 1\right)$.

\subsection{Co-data simulation}
\label{app:c1}

Here we give insights on how co-data information is simulated for the assessment of the Variational approximation accuracy with respect to the Gibbs sampler in Section \ref{subsec:GSvsVI} of the main article. Moreover we provide the graphical representation of the AUC for variable selection for both methods. 

The four degrees of co-data information are simulated as:
\begin{itemize}
    \item[] $G0\big)$ \textbf{no co-data} set-up: we include in the co-data matrix only the intercept, therefore $\mathbf{Z} = \mathbf{1}_p$;
    \item[] $G1\big)$ \textbf{non-informative} set-up: a binary co-data source is included in the model by randomly selecting $30$ regressors, therefore the co-data matrix $\mathbf{Z}$ is created from the binary vector $\mathbf{z} \in \left\{0, 1\right\}^p$, where $z_j = 1$ if the $j$-th variable is selected, $z_j = 0$ otherwise;
    \item[] $G2\big)$ \textbf{informative} set-up: a binary co-data source is included in the model by randomly selecting $20$ of the true non-zero regressors and $10$ of the true zero regressors, therefore the co-data matrix $\mathbf{Z}$ is created from the binary vector $\mathbf{z} \in \left\{0, 1\right\}^p$, where $z_j = 1$ if the $j$-th variable is selected, $z_j = 0$ otherwise;
    \item[] $G3\big)$ \textbf{perfect co-data information} set-up: the co-data matrix $\mathbf{Z}$ is created from the binary vector $\mathbf{z} \in \left\{0, 1\right\}^p$, where $z_j = 1$ if $\beta_j^0 \ne 0$, $z_j = 0$ otherwise.
\end{itemize}

\newpage
\section{Web Appendix G}
\subsection{Web Algorithm 1}

\begin{algorithm}[h!]
\small
        \caption{Gibbs sampler for Informative Horseshoe regression} 
 	{\textbf{Input}: $B, bn \in \mathbb{N}$, $\mathbf{y} \in \mathbb{R}^n$, $\mathbf{X} \in \mathbb{R}^{n \times (p+1)}$, $\mathbf{Z}_1, \dots, \mathbf{Z}_D \in \mathbb{R}^{p \times m_d}$, $\mathbf{a}, \mathbf{b} \in \mathbb{R}_+^D$, $s_0^2 \in \mathbb{R}_+$}\;
 	
 	\tcp{Set $\lambda_0 = 1$, $\boldsymbol{\lambda} = \mathbf{1}_p$, $\boldsymbol{\gamma} = \mathbf{0}_M$ and $\sigma^2, \tau^2 = 1$ and sample parameters $\boldsymbol{\varphi}$, $\boldsymbol{\kappa}^2$ and $\zeta$ from their prior distributions}
        
 	\For(){$b = 1 \text{ to } B$} 
 	{
 		sample $\boldsymbol{\beta} \mid \mathbf{y}, \mathbf{X}, \sigma^2, \tau^2, \lambda_0^2, \boldsymbol{\lambda} \sim \mathcal{N}_{p+1}\left(\boldsymbol{\Sigma}_\beta^\star \mathbf{X}^\intercal \mathbf{y}, \sigma^2 \boldsymbol{\Sigma}_\beta^\star \right)$, where $\boldsymbol{\Sigma}_\beta^\star = \left(\mathbf{X}^\intercal \mathbf{X} + \tau^{-2}\boldsymbol{\Lambda}^{-2}\right)^{-1}$\;
 		
 		sample $\lambda_0^2 \mid \beta_0, \tau^2, \sigma^2 \sim \mathcal{IG}\left(1, \frac{1}{\psi_0} + \frac{\beta_0^2}{2\sigma^2 \tau^2}\right)$\;
        
        sample $\psi_0 \mid \lambda_0^2 \sim \mathcal{IG}\left(1, 1 + \frac{1}{\lambda_0^2} \right)$\;
 		\For(){$j = 1 \text{ to } p$} 
 		{
 			sample $\lambda_j \mid \mathbf{Z}, \beta_j, \sigma^2, \tau^2, \boldsymbol{\gamma}, \varphi_j^2 \propto \lambda_j^{-1} e^{-\frac{\beta_j^2}{2\sigma^2\tau^2\lambda_j^2} - \frac{\lambda_j^2}{2 s_0^2\varphi_j^2} + \frac{\mu_j \lambda_j}{s_0^2 \varphi_j^2}} \cdot \mathbb{I}_{\left(\lambda_j > 0\right)}$ following the procedure in Section \ref{sec:rejection} of Supporting Information\;
 			sample $\varphi_j^2 \mid \mathbf{Z}, \lambda_j, \boldsymbol{\gamma} \sim \mathcal{IG}\left(1, \frac{1}{2} + \frac{(\lambda_j - \mu_j)^2}{2 s_0^2}\right)$\;
		}
 		sample $\boldsymbol{\gamma} \mid \mathbf{Z}, \boldsymbol{\lambda}, \boldsymbol{\varphi}, \boldsymbol{\kappa}^2 \sim \mathcal{N}_{M} \left(\boldsymbol{\Sigma}_{\gamma}^\star \left(\mathbf{Z}^\intercal \boldsymbol{\Phi}^{-2} \boldsymbol{\lambda}\right), s_0^2 \boldsymbol{\Sigma}_{\gamma}^\star\right)$, where $\boldsymbol{\Sigma}_{\gamma}^\star = \left( \mathbf{Z}^\intercal \boldsymbol{\Phi}^{-2} \mathbf{Z} + s_0^2\mathbf{D}_{\kappa}^{-1}\right)^{-1}$ and $\mathbf{D}_{\kappa} = \text{diag}\left(\kappa_1^2 \mathbf{1}_{m_1}, \dots, \kappa_D^2 \mathbf{1}_{m_D}\right)$\;
 		 \For(){$d = 1 \text{ to } D$} 
 		{
 			sample $\kappa_d^2 \mid \boldsymbol{\gamma}_d \sim \mathcal{IG} \left(a_d + \frac{m_d}{2}, b_d + \frac{\boldsymbol{\gamma}_d^\intercal\boldsymbol{\gamma}_d}{2}\right)$\;
		}
 		sample $\tau^2 \mid \boldsymbol{\beta}, \sigma^2, \lambda_0^2, \boldsymbol{\lambda}, \zeta \sim \mathcal{IG} \left(\frac{p}{2} + 1, \frac{1}{\zeta} + \frac{\beta_0^2}{2 \sigma^2\lambda_0^2} + \frac{1}{2\sigma^2}\sum_{j=1}^p \frac{\beta_j^2}{\lambda_j^2} \right)$\;
 		sample $\zeta \mid \tau^2 \sim \mathcal{IG}\left(1, 1 + \frac{1}{\tau^2}\right)$\;
 		sample $\sigma^2 \mid \mathbf{y}, \mathbf{X}, \boldsymbol{\beta}, \tau^2, \lambda_0^2, \boldsymbol{\lambda} \sim \mathcal{IG}\left(v + \frac{n+p+1}{2}, q + \frac{1}{2}\Vert \mathbf{y} - \mathbf{X} \boldsymbol{\beta}\Vert_2^2 + \frac{\beta_0^2}{2 \tau^2\lambda_0^2} + \frac{1}{2\tau^2}\sum_{j = 1}^p \frac{\beta_j^2}{\lambda_j^2}\right)$\;
 	 	\If() {$b > bn$} 
 	 	{
 	 		save $\boldsymbol{\beta}$, $\sigma^2$, $\tau^2$, $\lambda_0^2$, $\boldsymbol{\lambda}$ , $\boldsymbol{\gamma}$ and $\boldsymbol{\kappa}^2$
		}
 	}
 	\textbf{return} saved values\;
	\label{alg:algGS}
\end{algorithm}

\newpage
\subsection{Web Algorithm 2}
\begin{algorithm}[h!]
\small
	\label{alg:algVB}
	\caption{Variational Bayes approximation for informative Horseshoe regression} 
 	{\textbf{Input}: $\mathbf{y} \in \mathbb{R}^n$, $\mathbf{X} \in \mathbb{R}^{n \times (p+1)}$, $\mathbf{Z}_1, \dots, \mathbf{Z}_D \in \mathbb{R}^{p \times m_d}$, $v, q \in \mathbb{R}^+$, $\mathbf{a}, \mathbf{b} \in \mathbb{R}_+^D$, $s_0^2 \in \mathbb{R}_+$}\;
 	
 	\tcp{Set $b = 1$, $\epsilon = 10^{-3}$, $\mathcal{L}^{(0)} = -\infty$ and initialize all the needed moments}

 	\While(){$\mathcal{L}^{(b)} - \mathcal{L}^{(b-1)} > \epsilon$} 
 	{
 		Update parameter $\boldsymbol{\mu}_\beta^\star$ and compute the quantities $\text{diag}\left(\boldsymbol{\Sigma}_\beta^\star\right)$, $\vert\boldsymbol{\Sigma}_\beta^\star\vert$ and $\mathbf{X} \boldsymbol{\Sigma}_\beta^\star \mathbf{X}^\intercal$ as in Section \ref{app:b1} of Supporting Information\;
 		Update parameters $a_0^\star$ and $k_0^\star$\;
 		Update parameters $a_j^\star$, $b_j^\star$, $c_j^\star$ and $d_j^\star$ and evaluate the normalizing constant $s_j$, $\mathbb{E}_\lambda\left[\lambda_j\right]$, $\mathbb{E}_\lambda\left[\lambda_j^2\right]$ and $\mathbb{E}_\lambda\left[\lambda_j^{-2}\right]$ with numerical integration, for $j = 1, \dots, p$\;
 		Update parameters $\boldsymbol{\mu}_\gamma^\star$ and $\boldsymbol{\Sigma}_\gamma^\star$\;
 		Update parameters $e_d^\star$ and $ f_d^\star$, for $d = 1, \dots, D$\;
 		Update parameters $g^\star$ and $h^\star$\;
 		Update parameter $l^\star$\;
 	 	Compute $\mathcal{L}^{(b)}$ and set $b = b+1$\;
 	}
 	\textbf{return} $\boldsymbol{\mu}_\beta^\star$, $\boldsymbol{\Sigma}_\beta^\star$, $a_0^\star$, $\mathbf{a}^\star$, $\mathbf{b}^\star$, $\mathbf{c}^\star$, $\boldsymbol{\mu}_\gamma^\star$, $\boldsymbol{\Sigma}_\gamma^\star$, $\mathbf{e}^\star$, $\mathbf{f}^\star$, $g^\star$ and $l^\star$\;
\end{algorithm}

\newpage
\subsection{Web Algorithm 3}

\begin{algorithm}[h!]
\small
	\label{alg:algVBprobit}
	\caption{Variational Bayes approximation for probit informative Horseshoe regression} 
 	{\textbf{Input}: $\mathbf{y} \in \left\{0, 1\right\}^n$, $\mathbf{X} \in \mathbb{R}^{n \times (p+1)}$, $\mathbf{Z}_1, \dots, \mathbf{Z}_D \in \mathbb{R}^{p \times m_d}$, $\mathbf{a}, \mathbf{b} \in \mathbb{R}_+^D$, $s_0^2 \in \mathbb{R}_+$}\;
 	
 	\tcp{Set $b = 1$, $\epsilon = 10^{-3}$, $\mathcal{L}^{(0)} = -\infty$ and initialize all the needed moments}
 	
 	\While(){$\mathcal{L}^{(b)} - \mathcal{L}^{(b-1)} > \epsilon$} 
 	{
 		Update parameter $\mu_i^\star$, for $i = 1, \dots, n$\;
 		Update parameter $\boldsymbol{\mu}_\beta^\star$ and compute the quantities $\text{diag}\left(\boldsymbol{\Sigma}_\beta^\star\right)$ and $\vert\boldsymbol{\Sigma}_\beta^\star\vert$ as in Section \ref{app:b1} of Supporting Information\;
 		Update parameters $a_0^\star$ and $k_0^\star$\;
 		Update parameters $a_j^\star$, $b_j^\star$, $c_j^\star$ and $d_j^\star$ and evaluate the normalizing constant $s_j$, $\mathbb{E}_\lambda\left[\lambda_j\right]$, $\mathbb{E}_\lambda\left[\lambda_j^2\right]$ and $\mathbb{E}_\lambda\left[\lambda_j^{-2}\right]$ with numerical integration, for $j = 1, \dots, p$\;
 		Update parameters $\boldsymbol{\mu}_\gamma^\star$ and $\boldsymbol{\Sigma}_\gamma^\star$\;
 		Update parameters $e_d^\star$ and $ f_d^\star$, for $d = 1, \dots, D$\;
 		Update parameters $g^\star$ and $h^\star$\;
 	 	Compute $\mathcal{L}^{(b)}$ and set $b = b+1$\;
 	}
 	\textbf{return} $\boldsymbol{\mu}_\beta^\star$, $a_0^\star$,  $\mathbf{a}^\star$, $\mathbf{b}^\star$, $\mathbf{c}^\star$, $\boldsymbol{\mu}_\gamma^\star$, $\boldsymbol{\Sigma}_\gamma^\star$, $\mathbf{e}^\star$, $\mathbf{f}^\star$, $g^\star$\;
\end{algorithm}

\newpage
\section{Web Appendix H}

\subsection{Web Figure 1}

\begin{figure}[h!]
	\centering
	\includegraphics[scale = 0.45]{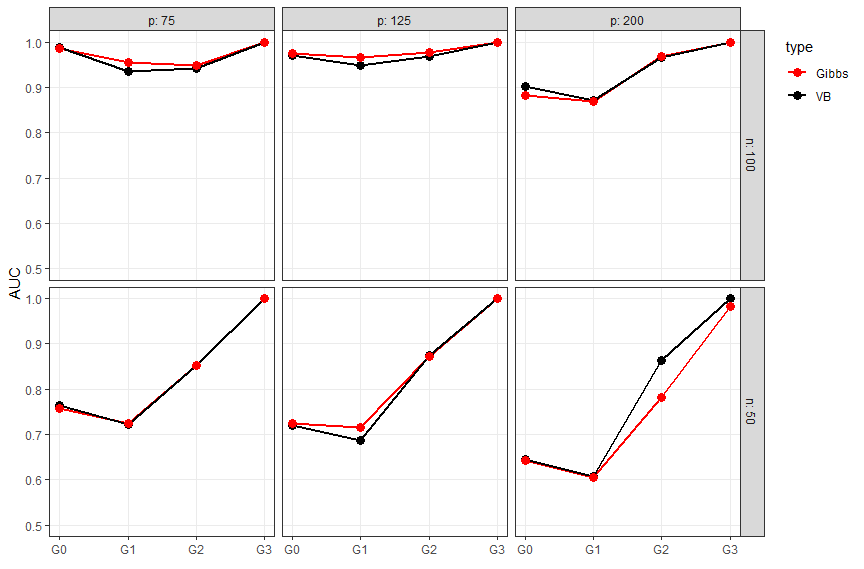}
	\caption{Variable selection with Gibbs sampler and Variational algorithm; the $AUC$ is averaged over $10$ replicates of each experiments.}
	\label{fig:sim1}
\end{figure}

\newpage
\subsection{Web Figure 2}

\begin{figure}[h!]
	\centering
	\includegraphics[scale = 0.5]{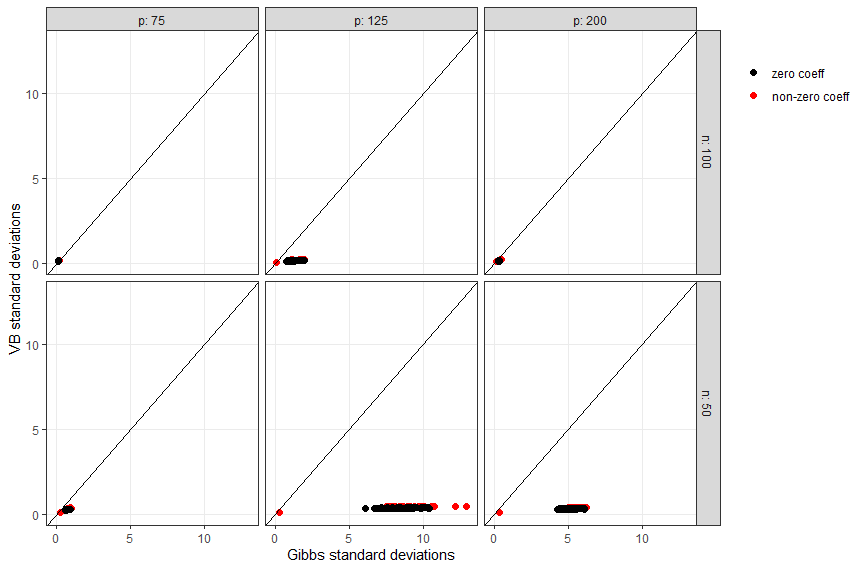}
	\caption{Estimated standard deviations of regression coefficients $\boldsymbol{\beta}$ with Gibbs sampler and Variational algorithm; results are evaluated over $10$ replicates of each experiments. Data have been aggregated for the different type of co-data.}
	\label{fig:sim_sd}
\end{figure}

\end{document}